\documentclass[12pt]{article}
\usepackage{amsmath}

\usepackage{algorithmic}
\usepackage{graphicx,psfrag,epsf}
\usepackage{enumerate}
\usepackage{natbib}
\usepackage{url} 
\usepackage{enumitem} 
\usepackage{multirow}
\usepackage{accents}
\usepackage{graphicx}
\usepackage{subcaption}
\usepackage{mwe}
\usepackage{float}
\usepackage{booktabs,makecell}
\usepackage{amsthm,amssymb}
\usepackage{amsfonts} 
\usepackage{dsfont}
\usepackage{amsmath}
\usepackage{amssymb}
\usepackage{setspace}
\usepackage{hyperref}
\usepackage{multirow}
\usepackage{booktabs}
\usepackage{bbm}
\usepackage{mathrsfs} 
\usepackage{bbm}
\usepackage{rotating}
\hypersetup{
     colorlinks   = true,
  linkcolor=blue,
            urlcolor=blue,
            citecolor=blue
}
\usepackage{xr}
\makeatletter
\newcommand*{\addFileDependency}[1]{
  \typeout{(#1)}
  \@addtofilelist{#1}
  \IfFileExists{#1}{}{\typeout{No file #1.}}
}
\makeatother

\newcommand*{\myexternaldocument}[1]{%
    \externaldocument{#1}%
    \addFileDependency{#1.tex}%
    \addFileDependency{#1.aux}%
}

\myexternaldocument{sup}
\usepackage[ruled,vlined,onelanguage]{algorithm2e}
\usepackage{pdflscape}
\usepackage{fancyhdr}

\fancypagestyle{mylandscape}{
\fancyhf{} 
\fancyfoot{
\makebox[\textwidth][r]{
  \rlap{\hspace{.75cm}
    \smash{
      \raisebox{4.87in}{
        \rotatebox{90}{\thepage}}}}}}
}

\newcommand{\blind}{0}
  
\begin{document}

\def\spacingset#1{\renewcommand{\baselinestretch}%
{#1}\small\normalsize} \spacingset{1}


\def\bmY{\boldsymbol{Y}}
\def\bmW{\boldsymbol{W}}
\def\bmM{\boldsymbol{M}}
\def\bmE{\boldsymbol{E}}
\def\bmL{\boldsymbol{L}}
\def\bmlambda{\boldsymbol{\lambda}}
\def\bmrho{\boldsymbol{\rho}}
\def\Cov{{\rm Cov}}
\def\Avar{{\rm Avar}}
\def\rmvec{{\rm vec}}
\def\bmSigma{\boldsymbol{\boldsymbol{\Sigma}}}
\def\mbR{\mathbb{R}}
\def\mbW{\mathbb{W}}
\def\bmI{\boldsymbol{I}}
\def\bmy{\boldsymbol{y}}
\def\bmvarepsilon{\boldsymbol{\varepsilon}}
\def\bmxi{\boldsymbol{\xi}}
\def\bmtheta{\boldsymbol{\theta}}
\def\hbmtheta{\boldsymbol{\widehat\theta}}
\def\tbmtheta{\boldsymbol{\widetilde\theta}}
\def\bbmtheta{\boldsymbol{\bar\theta}}
\def\mL{\mathscr{L}}
\def\tmL{\widetilde {\mL}}
\def\bmeta{{\boldsymbol{\eta}}}
\def\lbk{\left\{}
\def\rbk{\right\}}
\def\lak{\left|}
\def\rak{\right|}
\def\lmk{\left[}
\def\rmk{\right]}
\def\lsk{\left(}
\def\rsk{\right)}
\def\bmx{\boldsymbol{x}}
\def\bmz{\boldsymbol{z}}
\def\bmv{\boldsymbol{v}}
\def\bmmu{\boldsymbol{\mu}}
\def\bmzeta{\bmzeta}
\def\bmsigma{\boldsymbol{\sigma}}
\def\bmtheta{{\boldsymbol{\theta}}}
\def\bml{\boldsymbol{l}}
\def\bmo{\boldsymbol{0}}
\def\diag{{\rm diag}}
\def\bmrho{\boldsymbol{\rho}}
\def\E{{\rm E}}
\def\bmgamma{\boldsymbol{\gamma}}
\def\bmalpha{\boldsymbol{\alpha}}
\def\bmvtheta{\widehat{\bmtheta}}
\def\laak{\left\|}
\def\raak{\right\|}
\def\OLS{{\rm OLS}}
\def\GLS{{\rm GLS}}
\def\FGLS{{\rm FGLS}}
\def\LS{{\rm LS}}
\def\FLS{{\rm FLS}}
\def\tr{{\rm tr}}
\def\mbY{\mathbb{Y}}
\def\mbX{\mathbb{X}}
\def\rmE{{\rm E}}
\def\rmI{{\rm I}}
\def\bmI{{\boldsymbol{I}}}
\def\rmm{{\rm m}}
\def\rmP{{\rm P}}
\def\Cov{{\rm Cov}}
\def\var{{\rm Var}}
\def\mS{{\mathcal{S}}}
\def\mN{{\mathcal{N}}}
\def\rmU{{\mathrm{U}}}
\def\rmV{{\mathrm{V}}}
\def\rmQ{{\mathrm{Q}}}
\def\bmU{\boldsymbol{U}}
\def\bmV{{\boldsymbol{V}}}
\def\bmQ{{\boldsymbol{Q}}}
\def\bmH{{\boldsymbol{H}}}
\def\sgn{\operatorname{sgn}}
\def\bu{\boldsymbol u}
\def\mumed{\mu_{\text{med}}}
\def\vech{\mathrm{vech}}
\def\vec{\mathrm{vec}}
\def\bmZ{\boldsymbol{Z}}
\def\bmG{\boldsymbol{G}}
\def\bmX{\boldsymbol{X}}
\def\bmS{\boldsymbol{S}}
\def\bmh{\boldsymbol{h}}
\def\bmu{\boldsymbol{u}}
\def\bmzeta{\boldsymbol{\zeta}}
\def\bmt{\boldsymbol{t}}
\def\bmcalV{\boldsymbol{\mathcal{V}}}
\def\bms{\boldsymbol{s}}
\def\bmq{\boldsymbol{q}}
\def\bmk{\boldsymbol{k}}
\def\bmcalH{\boldsymbol{\mathcal{H}}}
\def\bmg{\boldsymbol{g}}
\def\bmw{\boldsymbol{w}}
\def\bmTheta{\boldsymbol{\Theta}}
\def\bmcalA{\boldsymbol{\mathcal{A}}}
\def\bmcalB{\boldsymbol{\mathcal{B}}}
\def\bmrmB{\boldsymbol{\mathrm{B}}}
\def\bmrmV{\boldsymbol{\mathrm{V}}}
\def\bmDelta{\boldsymbol{\Delta}}
\def\for{\textbf{\text{for }}}
\def\do{\textbf{\text{do }}}
\def\endfor{\textbf{\text{end for }}}
\def\mG{\mathcal{G}}
\def\bmcalG{\boldsymbol{\mathcal{G}}}

\def\be{\begin{equation}}
\def\ee{\end{equation}} 
\def\ben{\begin{equation*}}
\def\een{\end{equation*}}
\def\bea{\begin{eqnarray}}
\def\eea{\end{eqnarray}}
\def\bda{\begin{eqnarray*}}
\def\eda{\end{eqnarray*}}
\numberwithin{equation}{section}
\def\m{\color{magenta}}
\def\b{\color{blue}}
\def\red{\color{red}}
\def\here{{\color{magenta}Tao is up to here.}}
\def\bmY{\boldsymbol{Y}}
\def\bmX{\boldsymbol{X}}
\def\bmT{\boldsymbol{T}}
\def\bmA{\boldsymbol{A}}
\def\bmPsi{{\boldsymbol{\Psi}}}
\def\bmpsi{{\boldsymbol{\psi}}}
\def\bmUpsilon{\boldsymbol{\Upsilon}}
\def\bmcalZ{\boldsymbol{\mathcal{Z}}}
\def\bmrmZ{\boldsymbol{\mathrm{Z}}}
\def\bmbbE{\boldsymbol{\mathbb{E}}}
\def\bmmP{\boldsymbol{\mathcal{P}}}
\def\bmmH{\boldsymbol{\mathcal{H}}}
\def\bmvartheta{\boldsymbol{\bmtheta}}
\def\bmnu{\boldsymbol{\nu}}
\def\bmvarsigma{\boldsymbol{\varsigma}}
\def\bmvarSigma{\boldsymbol{\varSigma}}
\def\bmD{\boldsymbol{D}}

\newcommand{\Grad}{\nabla\!\!\!\!\nabla}
\newtheorem{theorem}{Theorem}
\newtheorem{remark}{Remark}
\newtheorem{proposition}{Proposition}
\newtheorem{corollary}{Corollary}
\newtheorem{assumption}{Assumption}
\newtheorem{lemma}{Lemma}
\newtheorem{definition}{Definition}
\newtheorem{example}{Example}


\if0\blind
{\spacingset{1.5} 
  \title{\bf   {On Robust Aggregation for Distributed Data}}
  \date{}
  \author{Xian Li, Xuan Liang, A. H. Welsh and Tao Zou 
 \\
    \textit{The Australian National University}}
  \maketitle

} \fi

\if1\blind
{\spacingset{2} 
  \bigskip
  \bigskip
  \bigskip
  \begin{center}
    {\LARGE\bf On Robust Aggregation for Distributed Data}
\end{center}
  \medskip
} \fi

\bigskip
\begin{abstract}
{

When data are stored across multiple locations, directly pooling all the data together for statistical analysis may be impossible due to communication costs and privacy concerns. Distributed computing systems allow the analysis of such data, by getting local servers to separately process their own statistical analyses and using a central processor to aggregate the local statistical results.
Naive aggregation of local statistics using simple or weighted averages, is vulnerable to contamination within a distributed computing system.
This paper develops and investigates a Huber-type aggregation method for locally computed $M$-estimators to handle contamination in the local estimates. 
Our implementation of this aggregation method requires estimating the asymptotic variance-covariance matrix of the $M$-estimator, which we accomplish using a robust spatial median approach. Theoretically, the Huber-type aggregation achieves the same convergence rate as if all the data were pooled.  We establish its asymptotic normality for making inferences, including justifying a two-step approach for detecting contamination in the distributed computing system.
Extensive simulation studies are conducted to validate the theoretical results and the usefulness of our proposed approach is demonstrated on  U.S. airline data.

}
\spacingset{1.2}

\end{abstract}

\noindent%
\vfill

\newpage
\spacingset{1.5} 

\section{ Introduction}
\label{sec:intro}

With accelerating developments in technology for gathering information, massive datasets have become increasingly pervasive (\citealp{fan2014challenges}).  For convenience in data collection or due to storage bottlenecks, large datasets are often spread across multiple machines (hereafter referred to simply as ``distributed data''), and it can be extremely difficult to gather all the data on a common machine.   The   obstacles include: communication costs, network bandwidth constraints (see, e.g., \citealp{nettleton2003decentralised},  \citealp{amirsadri2012computationally}), data ownership, privacy and security  (see, e.g., \citealp{duan2018odal}, \citealp{hu2022collaborative}).

Distributed computing has emerged as a popular tool for the analysis of datasets spread across multiple locations.  A common conceptualization of a distributed computing system is as  a single central processor connected to multiple local servers \citep{fan2019distributed}.
The local servers typically represent subsystems, each housing a portion of the full dataset, on which statistical data analysis can be conducted separately without any communication with other parts of the system. 
Through these data analyses, specific local statistics, such as estimates and standard errors, are produced by each local server, and then sent to the central processor for aggregation.  The central processor can act both as a local server, holding some data and producing its own local statistics,  and as the receiver of statistics from the other local servers for aggregation (see, e.g., \citealp{tu2021variance}).  The motivation for generating aggregated statistics at the central processor is that these potentially encompass information from the full dataset and hence should result in more accurate estimates or better statistical inferences than the separate local statistics. In contrast to gathering the data at the central processor for an analysis of the full data, aggregating local statistics reduces the communication costs and potentially enhances data privacy and security.  

 In recent years, there has been considerable literature developing methodology for distributed computing systems. For instance,
 \cite{li2013statistical} discussed the problem of density
estimation for distributed datasets; \cite{zhang2015divide} studied
distributed kernel ridge regression;  \cite{kleiner2014scalable} proposed
the bag of little bootstraps which  is well suited to distributed computing systems; both \cite{lee2017communication} and \cite{battey2018distributed}  investigated the distributed shrinkage
regression problem; and \cite{fan2019distributed} considered a distributed  algorithm for estimation of
eigenspaces.  All these works have a similar spirit of aggregating  or  averaging local statistics at a central processor within their respective application fields. 

Let $\widehat{\bmtheta}_{k}=(\widehat\theta_{k,1},\cdots,\widehat\theta_{k,p})^\top$ denote the local estimator of an unknown parameter vector of interest $\bmtheta=(\theta_1,\cdots,\theta_p)^\top$ from the $k$-th local server $k=1,\cdots,K$. %
Then  \citet{zhang2012communication} showed that the average  of $\widehat{\bmtheta}_{k}$ can estimate the parameter vector $\bmtheta$  at the rate of $1/\sqrt{N}$, where $N$ represents the size of the full sample, the best possible rate if the full data were gathered for estimation (referred to simply as full sample estimation). Furthermore, \citet{zhu2021least} showed that the weighted average can estimate $\bmtheta$ at the same rate $1/\sqrt{N}$, and also achieve the same asymptotic efficiency as full sample estimation.
These results hold when the $\widehat{\bmtheta}_{k}$ received at the central processor are uncontaminated; simple or weighted average aggregation of $\widehat{\bmtheta}_{k}$ can produce distorted results under contamination.

In a distributed system, contamination of local estimates $\widehat{\bmtheta}_{k}$ can occur in the data stored on one or more local servers, producing contaminated $\widehat{\bmtheta}_{k}$, or directly on the $\widehat{\bmtheta}_{k}$. For example, before  the $\widehat{\bmtheta}_{k}$ are transmitted from local servers,  they can be contaminated by (i) local server failure (due to being powered down for maintenance or by hardware errors, etc; see, e.g., \citealp{lamport1982byzantine}); or (ii) intentional contamination to protect data privacy and security (see, e.g., \citealp{teoh2006random}).  After transmission, the $\widehat{\bmtheta}_{k}$ can be contaminated by failures of the communication channels between local servers and the central processor, often due to network attacks or bandwidth constraints  (see, e.g.,  \citealp{xia2008attack} and \citealp{gill2011understanding}).
To reduce the effects of possible contamination in the local estimates, 
\cite{yin2018byzantine} used the elementwise median of $\widehat\theta_{k,j}$ across $k=1,\cdots,K$ to estimate $\theta_j$ for each dimension $j=1,\cdots, p$.  To improve the efficiency of the aggregated statistic,  \cite{tu2021variance} suggested using the average of several quantiles of $\widehat\theta_{k,j}$ across $k=1,\cdots,K$.  Other robust aggregations in the literature include 
marginal median (\citealp{xie2018generalized}), mean-around-median (\citealp{xie2018generalized}),  iterative filtering (\citealp{su2019securing,yin2019defending}), etc.

The aforementioned works successfully established some robust aggregation methods, but there are aspects that need to be strengthened to broaden the application of robust aggregation. 
First, these works only aggregate local sample means, which is restrictive in practice. 
Second, most of the methods aggregate  $\widehat\theta_{k,j}$ elementwise for each $j=1,\cdots,p$, ignoring the dependence structure among different elements $\widehat \theta_{k,1},\cdots,\widehat \theta_{k,p}$, and potentially losing efficiency. 
We increase flexibility by considering the  $\widehat{\bmtheta}_{k}$  to be local $M$-estimators. These estimators include the least-squares estimators and maximum likelihood estimators (MLEs), and allow for flexibility in the choice of estimator while handling diverse parameters of interest.  We handle possible contamination efficiently by using a Huber-type aggregation approach to robustly aggregate the vectors $\widehat\bmtheta_k$ of $M$-estimators from the $k = 1, \cdots, K$ local servers.  Specifically, we premultiply the local $M$-estimators by the negative half power  of their estimated asymptotic variance-covariance matrix to make the elements asymptotically independent, and then apply the elementwise Huber $M$-estimator. 
We derive the $\sqrt{N}$-consistency and the asymptotic normality of our proposed Huber-type aggregation when the number of the contaminated $\widehat\bmtheta_k$ among $\widehat\bmtheta_1,\cdots,\widehat\bmtheta_K$ is of order $o(\sqrt{K})$.  
Our Huber-type aggregation allows users to select their preferred tuning constant $c$ for different asymptotic relative efficiencies (AREs). 
For example,  the proposed Huber-type aggregation with 
$c=1.345$ and $c=1.5$ can achieve $95.0\%$ and $96.4\%$ ARE, which are better than the $2/\pi\approx63.7\%$ ARE obtained by  median aggregation \citep{yin2018byzantine}, and are comparable to the $3/\pi\approx95.5\%$ ARE obtained by  averaged quantile aggregation \citep{tu2021variance}. 


Our proposed Huber-type aggregation of $\widehat\bmtheta_k$ requires estimation of the $p\times p$ asymptotic variance-covariance matrix $\boldsymbol{\Sigma}$ of the  $M$-estimators $n_k^{1/2}\widehat\bmtheta_k$, where $n_k$ is the sample size on the $k$-th server. We can estimate $\boldsymbol{\Sigma}$ by a local variance estimate $\widehat{\boldsymbol{\Sigma}}_{k_0}$ from a local server $k_0$ that is known to transmit its estimates without contamination, as suggested in numerous papers (see, e.g., \citealp{tu2021byzantine} and \citealp{tu2021variance}).  However, \cite{el2020genuinely} pointed out that in a genuinely distributed computing system, it is unknown in advance whether a local estimate (including the variance estimate $\widehat{\boldsymbol{\Sigma}}_k$) at any server is  contaminated or not, so we prefer a robust aggregation of local variance estimates $\widehat{\boldsymbol{\Sigma}}_k$ across servers $k=1,\cdots,K$.  We propose a novel robust aggregation approach for local variance estimates, and  
show that the proposed spatial median aggregation of $\widehat{\boldsymbol{\Sigma}}_k$ \citep[adapted from][]{brown1983statistical} is guaranteed to be positive definite and converges at the rate $1/\sqrt{N}$. 
 Positive definiteness guarantees that the negative half power is well-defined, and the fast convergence rate ensures the quality of the Huber-type aggregation of  
 $\widehat\bmtheta_k$, enabling us to make more accurate asymptotic inferences (such as constructing confidence intervals and testing hypotheses) about the parameter vector  $\bmtheta$.  The spatial median aggregation of local variance estimates and the proposed Huber-type aggregation of local estimates, also allow us to detect the contaminated $\widehat\bmtheta_k$ and $\widehat{\boldsymbol{\Sigma}}_k$ among different servers $k=1,\cdots, K$.    To the best of our knowledge, the distributed computing literature has not yet considered employing robust methods for contamination detection  (see, e.g., \citealp{koushanfar2003line}, \citealp{ji2010distributed} and \citealp{ji2021statistics}). Furthermore, the detection of contamination in local 
$M$-estimates and local variance estimates remains predominantly unexplored.

The rest of this article is organized as follows.  Sections \ref{methodology} and \ref{spatial_med} introduce the proposed Huber-type aggregation of local $M$-estimators and the spatial median aggregation of local variance estimators, respectively, and provide an analysis of their theoretical properties. Building on these results, Section \ref{sec:detect}
 presents a two-step approach for sequentially detecting which local $M$- and variance estimates are contaminated.
Section \ref{numberstud} presents simulation studies to validate the theoretical results and Section \ref{realdata} provides a real data example  to show the usefulness of our proposed approaches.  Section \ref{conclude} concludes the article with a brief discussion.  Technical conditions 
 are presented in the Appendix. All technical lemmas,  theoretical proofs, additional simulation results, and additional tables are relegated to the supplementary material.

\section{ Huber-type Aggregation of Local $M$-Estimators}\label{methodology}
\subsection{ Preliminaries}

Recall that in a distributed system, we consider $k=1,\cdots, K$ local servers, and 
 assume that   $N$ independent and identically distributed ($i.i.d.$) data vectors $\bmZ_1, \cdots, \bmZ_N\stackrel{i.i.d.}\sim\bmZ$ are held on these $K$  local servers.  Specifically, let $\mathcal{S}=\{1, \cdots, N\}$ denote the observation indices of the full dataset $\{\bmZ_1, \cdots, \bmZ_N\}$, and let $\mathcal{S}_{k}\subset \mathcal{S}$ denote the indices of the observations held on the $k$-th server, so $\mathcal{S}=\cup_{k=1}^K \mathcal{S}_{k}$.  We assume that any two subsets $\mathcal{S}_{k_{1}} $ and $\mathcal{S}_{k_{2}}$ are disjoint for  $k_{1} \neq k_{2}$ and $k_1,k_2\in\{1,\cdots, K\}$. In addition, 
let $n_k=|\mathcal{S}_{k}|$ be the cardinality of $\mathcal{S}_{k}$, i.e., the local sample size at the $k$-th server, so $N=\sum_{k=1}^{K}n_{k}$.  

The full sample estimator of an  unknown parameter vector of interest  $\bmtheta=(\theta_1,\cdots,\theta_p)^\top$ is often considered the benchmark for distributed estimation approaches (see, e.g.,   \citealp{jordan2019communication} and \citealp{zhu2021least}).  The full sample $M$-estimator maximizes the criterion function
\be
\overline{M}(\bmtheta)=\frac{1}{N}\sum_{i=1}^Nm(\bmZ_i;\bmtheta),
\label{global:loss} 
\ee
where $m(\bmZ_i;\bmtheta)$ is a known smooth function. For instance, $m(\bmZ_i;\bmtheta)$ could represent the log-likelihood function of each observation $\bmZ_i$, so maximizing (\ref{global:loss}) results in the maximum likelihood estimator (MLE). Alternatively, if we set $m(\bmZ_i;\bmtheta) = -(Y_i - \bmX_i^\top \bmtheta)^2$, where $\bmZ_i = (Y_i, \bmX_i^\top)^\top$, $Y_i$ is the response variable, and $\bmX_i$ is the $p$-dimensional covariate vector, maximizing  (\ref{global:loss}) yields the least-squares estimator.



As discussed in the Introduction, there are obstacles to gathering the full data on one computer in the distributed system that prevent  the full sample criterion function (\ref{global:loss}) and the full sample $M$-estimator  being obtained in practice. However,
 the local $M$-estimator $\widehat\bmtheta_k$ at each server $k\in\{1,\cdots,K\}$,  which maximizes the local criterion function
 \begin{align}
 \overline{M}_k(\bmtheta)=\frac{1}{n_k}\sum_{i \in \mathcal{S}_{k}} m\lsk \bmZ_i;\bmtheta\rsk
 \label{localcriterion}
 \end{align} 
based on the local samples indexed in $\mathcal{S}_k$, can be obtained. Following the idea of least-squares approximation in \cite{wang2007unified} and \cite{zhu2021least}, we can approximate the full sample criterion function $\overline{M}(\bmtheta)$ in (\ref{global:loss})  by
 \be\label{eq:lsa}
 \overline{M}(\bmtheta)=\sum_{k=1}^K\frac{n_k}{N}\overline{M}_k(\bmtheta)\approx \sum_{k=1}^K \frac{n_k}{N}\lbk \overline{M}_k(\widehat{\bmtheta}_k)+\widetilde{M}_k(\bmtheta)\rbk=\sum_{k=1}^K \frac{n_k}{N}\overline{M}_k(\widehat{\bmtheta}_k)+\sum_{k=1}^K \frac{n_k}{N}\widetilde{M}_k(\bmtheta),
 \ee
where we define
$
\widetilde{M}_k(\bmtheta) = (\bmtheta-\widehat{\bmtheta}_{k})^{\top}  (-\bmU)(\bmtheta-\widehat{\bmtheta}_{k})/2,
$ with  $\bmU=-\rmE\{\nabla^2 m(\bmZ;\bmtheta^{(0)})\}=  -\rmE\{\partial^2m(\bmZ;\bmtheta)/(\partial\bmtheta\partial\bmtheta^\top)|_{\bmtheta=\bmtheta^{(0)}}\}\in\mathbb{R}^{p\times p}$ and $\bmtheta^{(0)}$ is the true parameter vector of $\bmtheta$. To see why $ \overline{M}_k(\widehat{\bmtheta}_k)+\widetilde{M}_k(\bmtheta)$ is a good approximation to $\overline{M}_k(\bmtheta)$, suppose  $\bmtheta$ is near $\widehat{\bmtheta}_k$ and consider the 
Taylor series expansion of $\overline{M}_k(\bmtheta)$ at $\widehat{\bmtheta}_{k}$, which leads to
\bda
\overline{M}_k(\bmtheta)&=& \overline{M}_k(\widehat{\bmtheta}_k)+\frac{1}{2}\left(\bmtheta-\widehat{\bmtheta}_{k}\right)^{\top}\lbk{\nabla^2 \overline{M}_k(\widehat{\bmtheta}_k)}\rbk\left(\bmtheta-\widehat{\bmtheta}_{k}\right)+ o_{\rmP}\lsk\laak\bmtheta-\widehat{\bmtheta}_k\raak^2_2\rsk\nonumber\\
& =& \overline{M}_k(\widehat{\bmtheta}_k)+\widetilde{M}_k(\bmtheta)+ o_{\rmP}\lsk\laak\bmtheta-\widehat{\bmtheta}_k\raak^2_2\rsk\nonumber,
\eda
where $\laak\cdot\raak_2$ is the vector 2-norm, and $\nabla^2\overline{M}_k(\widehat{\bmtheta}_k)=n_k^{-1}\sum_{i \in \mathcal{S}_{k}}\nabla^2 m(\bmZ_i;\widehat{\bmtheta}_k)$.  The first equality in the above equation is due to the fact that the local $M$-estimator $\widehat{\bmtheta}_{k}$ solves  $\nabla \overline{M}_k (\bmtheta)=\partial \overline{M}_k (\bmtheta)/\partial \bmtheta=\boldsymbol{0}_p$, where  $\boldsymbol{0}_p$ is the $p$-dimensional vector of zeros; and the second equality follows from the consistency of the local $M$-estimator $\widehat{\bmtheta}_{k}$ as $n_k\to\infty$ under Conditions (C1) -- (C4) in the Appendix (e.g., implied by Lemma \ref{expansion} of the supplementary material) 
from which $\nabla^2\overline{M}_k(\widehat{\bmtheta}_k)$ converges in probability to $-\bmU$.

Since $\sum_{k=1}^K ({n_k}/{N})\overline{M}_k(\widehat{\bmtheta}_k)$ in (\ref{eq:lsa}) does not depend on $\bmtheta$, maximizing the full sample criterion function $\overline{M}(\bmtheta)$ can be approximated by maximizing $\sum_{k=1}^K ({n_k}/{N})\widetilde{M}_k(\bmtheta)$. Next, maximizing $\sum_{k=1}^K ({n_k}/{N})$ $\widetilde{M}_k(\bmtheta)$ can be achieved by solving the estimating equations $\sum_{k=1}^K ({n_k}/{N})$ $\partial \widetilde{M}_k(\bmtheta)/\partial \bmtheta=
\sum_{k=1}^K ({n_k}/{N}) (-\bmU)(\bmtheta-\widehat{\bmtheta}_{k})=\boldsymbol{0}_p,
$  
which can be further reduced to 
\begin{equation}
\sum_{k=1}^K \frac{n_k}{N}  \lsk\widehat{\bmtheta}_{k}-\bmtheta\rsk=\boldsymbol{0}_p,\label{LSA:0}
\end{equation}
provided that  the matrix $\bmU$ is non-singular as assumed  in Condition (C3) of the Appendix. The solution of the estimating equations (\ref{LSA:0}) is given by $\overline{\bmtheta}=\sum_{k=1}^K(n_k/N)\widehat{\bmtheta}_k$, i.e., the weighted average of the local $M$-estimators $\widehat{\bmtheta}_k$.  \cite{zhu2021least} has shown that a more general but similar weighted average aggregation can both estimate $\bmtheta^{(0)}$ at the rate of $1/\sqrt{N}$ and also achieve the same asymptotic efficiency as the full sample $M$-estimator that maximizes  (\ref{global:loss}). However,  the resulting aggregated estimator is not resistant to contamination of the local estimates $\widehat{\bmtheta}_{k}$ which can occur at  various stages in the distributed system; see the Introduction for details.  To achieve resistance to contamination, we introduce Huber-type robust aggregation of local $M$-estimators in the following subsection.

\subsection{ Huber-type Aggregation and Its Asymptotic Properties }\label{sec:RAEDD}

As discussed in the Introduction, the local $M$-estimates received at  the central processor could potentially be contaminated. To distinguish the uncontaminated local $M$-estimate $\widehat{\bmtheta}_k$ which maximizes the local criterion function $ \overline{M}_k(\bmtheta)$ in (\ref{localcriterion}),  we denote  the actual estimate that is received at  the central processor by   $\widehat{\bmtheta}_{k}^\star$
for $k=1,\cdots, K$. Accordingly, if $\widehat{\bmtheta}_{k}^\star\neq\widehat{\bmtheta}_{k}$, 
$\widehat{\bmtheta}_{k}^\star$ is contaminated; otherwise, $\widehat{\bmtheta}_{k}^\star$ is not contaminated.   In real applications, only the $\widehat{\bmtheta}_{k}^\star$s are  observed at the central processor,  
so the aggregation methods will be based on $\widehat{\bmtheta}_{k}^\star$.

To bound the influence of contaminated $\widehat{\bmtheta}_{k}^\star$ in the aggregation, note that under Conditions (C1) -- (C4) in the Appendix, we have 
\be\label{local assum}
\sqrt{n_{k}}\lsk\widehat{\bmtheta}_{k}-\bmtheta^{(0)}\rsk\stackrel{d}{\longrightarrow}\mathcal{N}\lsk\boldsymbol {0}_p,\bmSigma\rsk,\textrm{ as $n_k\to\infty$},
\ee
where $\boldsymbol{\Sigma}=\bmU^{-1}\bmV\bmU^{-1}$ with
$\bmV=\var\{
\nabla m(\bmZ;\bmtheta^{(0)})\}\in\mathbb{R}^{p\times p}$ the variance-covariance matrix of $\nabla m(\bmZ;\bmtheta^{(0)})$; see 
Lemma \ref{expansion}
   of   the supplementary material. 
Accordingly, 
$\boldsymbol{\Sigma}$ is the positive definite (by Condition (C3) in the Appendix) $p\times p$ asymptotic variance-covariance matrix of the  $M$-estimator $\widehat{\bmtheta}_{k}$.  In practice, $\bmSigma$ is unknown, so we will estimate it  with a consistent estimator  $\widehat{\boldsymbol{\Sigma}}$; how to obtain a  consistent estimator  $\widehat{\boldsymbol{\Sigma}}$ within the distributed system is discussed in (\ref{local_estimate_sigma}) and Section \ref{spatial_med} below.  For now, suppose that $\widehat{\boldsymbol{\Sigma}}$ is ready at the central processor. Then 
we have 
$\widehat{\bmSigma}^{-1/2}\hspace{-0.5em}\sqrt{n_{k}}(\widehat{\bmtheta}_{k}-\bmtheta^{(0)})\stackrel{d}{\longrightarrow}\mathcal{N}(\boldsymbol {0}_p,\boldsymbol {I}_p)$, where $\boldsymbol {I}_p$ is the $p$-dimensional identity matrix.  That is, the different elements in $\widehat{\bmSigma}^{-1/2}\hspace{-0.5em}\sqrt{n_{k}}(\widehat{\bmtheta}_{k}-\bmtheta^{(0)})$ are asymptotically independent and each is asymptotically $\mathcal{N}(0,1)$, justifying our treating all elements the same.  
We rewrite the estimating equations (\ref{LSA:0}) as $\sum_{k=1}^K ({n_k}/{N}) n_k^{-1/2} \{\widehat{\bmSigma}^{-1/2}\hspace{-0.5em} \sqrt{n_k}(\widehat{\bmtheta}_{k}-\bmtheta)\}=\boldsymbol{0}_p$, and then replace (\ref{LSA:0}) by the new estimating equations
\be\label{robust:estimating eq2}
\sum_{k=1}^K\frac{n_k}{N}\frac{1}{\sqrt{n_k}}{\bmpsi}_c\lbk\widehat{\boldsymbol{\Sigma}}^{-1/2}\sqrt{n_k}\lsk\widehat{\bmtheta}_{k}^\star-\bmtheta\rsk\rbk=\boldsymbol{0}_p,
\ee 
where ${\bmpsi}_c(\boldsymbol u)=(\psi_c(u_1),\cdots,\psi_c(u_p))^\top$ for $\bmu=(u_1,\cdots, u_p)^\top$, and $\psi_c(u)=
-c \mathbbm{1}{\lbk   u<-c\rbk}+
u\mathbbm{1}{\lbk   |u|\leq c\rbk}+
c  \mathbbm{1}{\lbk u>c\rbk}
$ for $u\in\mathbb{R}$ with  $\mathbbm{1}{\{\cdot\}}$ the indicator function and $c$ a positive tuning constant, is the Huber function (\citealp{huber1964robust}). 

The tuning constant $c$ restricts the influence of contaminated $\widehat{\bmtheta}_{k}^\star$ in the estimating equations; 
if  $c=\infty$, (\ref{robust:estimating eq2}) reduces to (\ref{LSA:0}), 
the solution of which is the  weighted average $\overline{\bmtheta}=\sum_{k=1}^K(n_k/N)\widehat{\bmtheta}_k^\star$. 
In real applications, the tuning constant $c$ can be chosen flexibly based on each user's preferred  asymptotic relative efficiency (ARE) of the Huber-type aggregation $\widehat\bmtheta$, where  $\widehat\bmtheta$ solves the estimating equations (\ref{robust:estimating eq2}). To establish the ARE of $\widehat\bmtheta$ relative to  the  weighted average aggregation $\overline\bmtheta$, we consider the asymptotic setting
\be\label{eq:asysetting}
\textrm{$K=K(N)\to\infty$ and $n=\frac{N}{K}=\frac{N}{K(N)}\to\infty$ as $N\to\infty$,  and 
$c_1\leq \frac{n_k}{n}\leq c_2$} 
\ee
  for all $k=1,\cdots,K$, where $c_1$ and $c_2$ are some finite positive constants and hence $n_k\asymp n$ uniformly in $k$. The same asymptotic setting is considered 
  in Section 2.1 of     \cite{zhu2021least}.  

Let $\mathscr{E}_\bmtheta=\{k: \widehat{\bmtheta}_{k}^\star\neq\widehat{\bmtheta}_{k}\}\subset\{1,\cdots,K\}$ be the set containing the indices of the contaminated $\widehat{\bmtheta}_{k}^\star$, so the cardinality $|\mathscr{E}_\bmtheta|$ of the set $\mathscr{E}_\bmtheta$ is the number of contaminated $\widehat\bmtheta_k^\star$. 
The $\sqrt{N}$-consistency and the asymptotic normality of the Huber-type aggregation $\widehat\bmtheta$ are established in the theorem given below.  
\begin{theorem}\label{uni:consist2}
Suppose  Conditions  (C1) -- (C5)  in  the Appendix and the Asymptotic Setting (\ref{eq:asysetting}) are satisfied. In addition, assume $n^2/{N}\to\infty$ and $|\mathscr{E}_\bmtheta|/\sqrt{K}\to 0$ as $N\to\infty$. If  $\widehat{\boldsymbol{\Sigma}}$ used in estimating equations (\ref{robust:estimating eq2}) satisfies  $\widehat{\boldsymbol{\Sigma}}-\bmSigma=O_\rmP(1/\sqrt{n})$, we have
\[
\sqrt{N}\lsk{ \widehat{\bmtheta}}-\bmtheta^{(0)}\rsk \stackrel{d}\longrightarrow \mathcal{N}\lsk\boldsymbol{0}_p, \tau_c^{-1}\bmSigma\rsk, 
\]
where  $\tau_c=b_c^{2} / \sigma_{c}^{2}$,  $b_c=\int_{-c}^{c} \mathrm{d} \Phi(u)$, $\sigma_{c}^{2} =\int_{-c}^{c} \psi_c^{2}(u) \mathrm{d} \Phi(u)+c^{2}(1-b_c)$,  and $\Phi(u)$ is the cumulative distribution function of  the standard normal distribution $\mathcal{N}(0,1)$. 
\end{theorem}

\noindent Conditions (C1) -- (C5), are  standard conditions for establishing the
asymptotic properties of $M$-estimators. 


If a local $M$-estimator $\widehat{\bmtheta}_{k}^\star$ is not contaminated, i.e., $\widehat{\bmtheta}_{k}^\star = \widehat{\bmtheta}_{k}$, then under the Asymptotic Setting (\ref{eq:asysetting}), it typically exhibits a bias of order $O(n^{-1})$ (see, e.g., Lemma B.6 in \citealp{huang2019distributed} and Proposition 1 in \citealp{zhu2021least}) and a variance of order $O(n^{-1})$ (implied by (\ref{local assum})). In distributed computing systems, it is commonly observed that aggregating $\widehat{\bmtheta}_{k}$ can reduce the variance to the order of $O\big((nK)^{-1}\big) = O(N^{-1})$. However, this aggregation does not improve the order of the bias, which remains at $O(n^{-1})$. As a consequence, the aggregation of $\widehat{\bmtheta}_{k}$ has a mean squared error (MSE) of order $O(N^{-1} + n^{-2}) = O(N^{-1})$ only when $n^2/N \to \infty$, meaning that the bias is negligible compared to the variance in this regime; see, for example, \citet{fan2019distributed}. This is also the intuitive reason why the Huber-type aggregation $\widehat{\bmtheta}$ can achieve the $\sqrt{N}$-consistency and asymptotic normality in Theorem \ref{uni:consist2} under the condition $n^2/{N}\to\infty$. Under the Asymptotic Setting (\ref{eq:asysetting}), $n^2/{N}\to\infty$ is equivalent to  $n/K\to\infty$, i.e., a large $n$ and small $K$ setting, which is common for distributed data. 


The set $\mathscr{E}_\bmtheta=\{k: \widehat{\bmtheta}_{k}^\star\neq\widehat{\bmtheta}_{k}\}$ in Theorem \ref{uni:consist2} is unknown in practice, but
solving estimating equations (\ref{robust:estimating eq2}) to obtain the  Huber-type aggregation $\widehat{\bmtheta}$ does not require us to know $\mathscr{E}_\bmtheta$. However, to ensure the $\sqrt N$-consistency and asymptotic normality of  $\widehat{\bmtheta}$, 
Theorem \ref{uni:consist2} needs the condition $|\mathscr{E}_\bmtheta|/\sqrt{K}\to 0$, namely the number of the contaminated $\widehat\bmtheta_k^\star$ must be of order $o(\sqrt{K})$. A similar  condition is also imposed in the robust distributed computing literature in order to guarantee asymptotic normality; see, e.g., an equivalent condition in Theorem 1  in \cite{tu2021variance}. 
 Such a requirement on the amount of contamination in  a distributed computing system is common and is often referred to as ``fault tolerance" (see \citealp{jalote1994fault}). 
To understand why the  Huber-type aggregation $\widehat{\bmtheta}$ is resistant to contamination, i.e., why $\widehat{\bmtheta}$ is $\sqrt N$-consistent and asymptotically normal under $|\mathscr{E}_\bmtheta|/\sqrt{K}\to 0$ as demonstrated in Theorem \ref{uni:consist2}, we examine part of the proof of Theorem \ref{uni:consist2} below. Specifically, to derive the asymptotic normality of $\widehat{\bmtheta}$ that solves the estimating equations (\ref{robust:estimating eq2}), the central limit theorem (CLT) needs to be developed for 
\bea
&&\sum_{k=1}^K\frac{n_k}{\sqrt N}\frac{1}{\sqrt{n_k}}{\bmpsi}_c\lbk\boldsymbol{\Sigma}^{-1/2}\sqrt{n_k}\lsk\widehat{\bmtheta}_{k}^\star-\bmtheta^{(0)}\rsk\rbk\nonumber\\
&\stackrel{\textrm{(i)}}=&\sum_{k\in \mathscr{E}_\bmtheta}\frac{n_k}{\sqrt N}\frac{1}{\sqrt{n_k}}{\bmpsi}_c\lbk\boldsymbol{\Sigma}^{-1/2}\sqrt{n_k}\lsk\widehat{\bmtheta}_{k}^\star-\bmtheta^{(0)}\rsk\rbk+\sum_{k\notin \mathscr{E}_\bmtheta}\frac{n_k}{\sqrt N}\frac{1}{\sqrt{n_k}}{\bmpsi}_c\lbk\boldsymbol{\Sigma}^{-1/2}\sqrt{n_k}\lsk\widehat{\bmtheta}_{k}-\bmtheta^{(0)}\rsk\rbk\nonumber\\
&\stackrel{\textrm{(ii)}}=&O_{\rmP}\lsk \frac{|\mathscr{E}_\bmtheta|}{\sqrt{K}}\rsk+\sum_{k\notin \mathscr{E}_\bmtheta}\frac{n_k}{\sqrt N}\frac{1}{\sqrt{n_k}}{\bmpsi}_c\lbk\boldsymbol{\Sigma}^{-1/2}\sqrt{n_k}\lsk\widehat{\bmtheta}_{k}-\bmtheta^{(0)}\rsk\rbk,\label{eq:CLTfore}
\eea
where equality (i) holds because $\widehat{\bmtheta}_{k}^\star=\widehat{\bmtheta}_{k}$ for $k\notin \mathscr{E}_\bmtheta$ are not contaminated, and  equality (ii) because of the boundedness of the vector-valued Huber function $\bmpsi_c(\cdot)$ and the Asymptotic Settings  (\ref{eq:asysetting}). Under $|\mathscr{E}_\bmtheta|/\sqrt{K}\to 0$, the first term $O_\rmP(|\mathscr{E}_\bmtheta|/\sqrt{K})=o_\rmP(1)$ is negligible and the CLT  for the second term in  (\ref{eq:CLTfore}) follows from the Lindeberg-Feller CLT; similar techniques are employed in the proofs of Theorem \ref{uni:consist} and Theorem \ref{uni:consist2} in Section \ref{proof:tm} of the supplementary material. The $\sqrt{N}$-consistency and the asymptotic normality of $\widehat{\bmtheta}$ require the condition on $|\mathscr{E}_\bmtheta|$ but impose no further restrictions on the contaminated estimates $\{\widehat{\bmtheta}_k^\star : k \in \mathscr{E}_\bmtheta\}$. In particular, the values of $\widehat{\bmtheta}_k^\star$ for $k \in \mathscr{E}_\bmtheta$ can be arbitrary; see Section \ref{numberstud}  for examples of different types of contaminated $\widehat{\bmtheta}_k^\star$.

From Theorem {\ref{uni:consist2}}, 
the asymptotic variance-covariance matrix of the Huber-type aggregation $\widehat{\bmtheta}$ is $\tau_c^{-1}\bmSigma$. From \cite{zhu2021least}, the (non-robust) weighted average aggregation $\overline{\bmtheta}$ satisfies $\sqrt{N}({\overline{\bmtheta}}-\bmtheta^{(0)}) \stackrel{d}\longrightarrow \mathcal{N}(\boldsymbol{0}_p, \bmSigma)$.   According to the definition of ARE as the ratio of the determinants of two estimators' asymptotic variance-covariance matrices (see, e.g., equation (3) in \citealp{serfling2011asymptotic}), we obtain that  $\tau_c$ is the ARE of $\widehat{\bmtheta}$ relative to  $\overline{\bmtheta}$. Based on the expression for $\tau_c$ given in  Theorem {\ref{uni:consist2}}, $\tau_c\uparrow 100\%$ as $c \to\infty$ (which is consistent with the fact that the Huber-type aggregation $\widehat{\bmtheta}$ reduces to the weighted average aggregation $\overline{\bmtheta}$ when $c=\infty$) and $\tau_c\downarrow (2/\pi)\approx63.7\%$ as $c\to 0$. 
In the classical robust statistics literature,  $c=1.345$ is used as the default value in the robust linear regression function \texttt{rlm} of the R package \texttt{MASS}. 
Other values of $c$ include: $c=0.9818$ ($\tau_c=90.0\%$), $c=1.2$ ($\tau_c=93.3\%$; see \citealp{cantoni2001robust}) and $c=1.25$ ($\tau_c=93.9\%$; see \citealp{chi1994m} and \citealp{street1988note}). The classical robust statistics literature focuses on contamination in the data, and the ARE $\tau_c$ is derived under the assumption that the uncontaminated data follow a Gaussian distribution. In contrast, the Huber-type aggregation $\widehat{\bmtheta}$ addresses the contamination of $\{\widehat{\bmtheta}_k^\star:k\in\mathscr{E}_\bmtheta\}$ received at the central processor, where contamination can occur at any stages within the distributed system, including contamination of data stored on local servers. In addition, the derivation of $\tau_c$ in Theorem \ref{uni:consist2} uses the asymptotic normality of the uncontaminated local estimators and
 does not require the data $\bmZ_1, \cdots, \bmZ_N$ to be Gaussian. In particular, the distribution of $\bmZ_1, \cdots, \bmZ_N$ can be arbitrary as long as it satisfies the standard conditions for establishing the asymptotic properties of $M$-estimators, i.e., Conditions (C1) -- (C5). For example, when $\bmZ_i = (Y_i, \bmX_i^\top)^\top$, $Y_i$ can be a binary response variable and we can aggregate local MLEs $\widehat{\bmtheta}_k^\star$, as in the numerical studies of binary regression discussed in Sections \ref{numberstud} and \ref{realdata} below.

\section{ Aggregation of Local Variance Estimators}\label{spatial_med}

To ensure the $\sqrt{N}$-consistency and asymptotic normality of $\widehat{\bmtheta}$, we require a positive definite, robust estimator $\widehat{\boldsymbol{\Sigma}}$ that satisfies the rate condition $\widehat{\boldsymbol{\Sigma}} - \bmSigma = O_\rmP(1/\sqrt{n})$ as specified in Theorem \ref{uni:consist2}. Positive definiteness is needed because: (i) implementing the estimating equations (\ref{robust:estimating eq2}) to obtain the Huber-type aggregation involves $\widehat\bmSigma^{-1/2}$, and positive definiteness naturally guarantees that $\widehat\bmSigma^{-1/2}$ is well-defined; and (ii) based on Theorem \ref{uni:consist2}, the asymptotic variance-covariance matrix of the Huber-type aggregation is $\tau_c^{-1}\bmSigma$, and using its estimator $\tau_c^{-1}\widehat\bmSigma$ and asymptotic normality to make inferences regarding $\bmtheta$, such as constructing confidence intervals and testing hypotheses, requires $\widehat\bmSigma$ to be positive definite.  We next propose approaches to obtain $\widehat{\boldsymbol{\Sigma}}$ that satisfies this condition.

If the data $\{\bmZ_i:i\in\mathcal{S}_k\}$ and $\widehat\bmtheta_k$ from local server $k$ are not contaminated, the asymptotic variance-covariance matrix $\boldsymbol{\Sigma}=\bmU^{-1}\bmV\bmU^{-1}$ can be estimated by the empirical local variance estimator 
\be
  \label{local_estimate_sigma}
    \begin{matrix}
    \widehat{\boldsymbol{\Sigma}}_{k} =\widehat{\bmU}_{k}^{-1}\widehat{\bmV}_{k}\widehat{\bmU}_{k}^{-1},
    \end{matrix}
  \ee 
  where   $\widehat{\bmV}_{k}={n_{k}^{-1}}\sum_{i\in\mathcal{S}_{k}}\{ \nabla m(\bmZ_i;\widehat{\bmtheta}_{k})-
  \nabla \overline{M}_k(\widehat{\bmtheta}_{k})\}\{\nabla m(\bmZ_i;\widehat{\bmtheta}_{k}) -
  \nabla\overline{M}_k(\widehat{\bmtheta}_{k})$ $\}^\top$ and $\widehat{\bmU}_{k}=-\nabla^2 \overline{M}_k(\widehat{\bmtheta}_{k})$ for $k=1,\cdots, K$. Using similar techniques to those used in the proof of Lemma \ref{expansion} in the supplementary material, 
  we can show that under Conditions (C1) -- (C4) and the Asymptotic Setting (\ref{eq:asysetting}), we have $\widehat{\boldsymbol{\Sigma}}_k - \bmSigma = O_\rmP(1/\sqrt{n_k})=O_\rmP(1/\sqrt{n})$.

Suppose that the local servers can transmit their local variance estimates to the central processor, but those estimates received at the central processor, denoted by $\widehat{\boldsymbol{\Sigma}}_k^\star$, could potentially be contaminated. Specifically, we collect all the indices of those contaminated $\widehat{\boldsymbol{\Sigma}}_k^\star$ in a set $\mathscr{E}_{\bmSigma}=\{k:\widehat{\boldsymbol{\Sigma}}_k^\star\neq \widehat{\boldsymbol{\Sigma}}_k\}\subset\{1,\cdots,K\}$ that is unknown in practice. Note that $\mathscr{E}_{\bmSigma}$ can be equal or not equal to $\mathscr{E}_\bmtheta=\{k: \widehat{\bmtheta}_{k}^\star\neq\widehat{\bmtheta}_{k}\}$ because the contamination of local variance estimates $\widehat{\boldsymbol{\Sigma}}_k^\star$ can also occur at different stages in the distributed system. 

If we can identify an uncontaminated local server in advance, i.e., find a $k_0$ such that $k_0\not\in \mathscr{E}_{\bmSigma}$, then the central processor can use its local variance estimator $\widehat{\boldsymbol{\Sigma}}_{k_0}$ as $\widehat{\boldsymbol{\Sigma}}$. While this approach 
is suggested in numerous papers (see, e.g., \citealp{tu2021byzantine} and \citealp{tu2021variance}), it suffers from the fact that $ \mathscr{E}_{\bmSigma}$ is not usually known in advance. This motivates us to consider a robust aggregation of the local variance estimates.

To the best of our knowledge, the aggregation of local variance estimators that satisfies the aforementioned three requirements remains largely unexplored in the robust distributed computing literature. 
Since the Huber-type aggregation introduced in Section \ref{sec:RAEDD} can achieve robustness and $\sqrt N$-consistency, we may consider using this approach  to aggregate $\widehat{\bmSigma}_k^\star$. Specifically, 
in order to obtain the Huber-type aggregation of $\widehat{\bmSigma}_k^\star$, we can let $\widehat{\bmtheta}_k^\star=\vech(\widehat{\bmSigma}_k^\star)$ in estimating equations (\ref{robust:estimating eq2}), 
where $\vech(\widehat{\bmSigma}_k^\star)$ is a ${p(p+1)/2}$-dimensional vector and the operator $\vech(\cdot)$ stacks the elements on and below the main diagonal of a generic symmetric matrix  (see, e.g., \citealp[p.662]{lutkepohl2005new}, for the definition of $\vech(\cdot)$). However, there are problems with directly implementing this approach. First, using the estimating equations (\ref{robust:estimating eq2}) with $\widehat{\bmtheta}_k^\star = \vech(\widehat{\bmSigma}_k^\star)$ involves estimation of the asymptotic variance-covariance matrix of  $\vech(\widehat{\bmSigma}_k)$. Since $\widehat{\bmSigma}_k$ in (\ref{local_estimate_sigma}) is already a variance estimator, deriving the asymptotic variance-covariance matrix for $\vech(\widehat{\bmSigma}_k)$ is tedious and requires higher-order moments of the derivatives of $m(\bmZ; \bmtheta)$ with respect to $\bmtheta$. Second, directly solving the estimating equations (\ref{robust:estimating eq2}) with $\widehat{\bmtheta}_k^\star = \vech(\widehat{\bmSigma}_k^\star)$ does not necessarily produce a positive definite estimator of $\bmSigma$. Third, the robustness and $\sqrt N$-consistency of the Huber-type aggregation in Theorem \ref{uni:consist2} is  established for $M$-estimators, but  $\widehat{\bmSigma}_k$ in (\ref{local_estimate_sigma}) is not directly an $M$-estimator. 
Finally, the  Huber-type aggregation is designed to achieve high asymptotic relative efficiency, which is less important than strong robustness in the aggregation of $\widehat{\bmSigma}_k^\star$. 

We can also consider using \citeauthor{yin2018byzantine}'s (\citeyear{yin2018byzantine}) median aggregation approach to aggregate each element of $\vech(\widehat{\bmSigma}_k^\star)$ across $k=1,\cdots,K$. However, like the Huber-type aggregation, the resulting estimator of $\bmSigma$ is not necessarily positive definite. In the following subsections, we propose the spatial median aggregation of $\widehat{\bmSigma}_k^\star$, which not only guarantees positive definiteness but is also robust to contamination and achieves a better convergence rate of $1/\sqrt{N}$. Since $\vech(\widehat{\bmSigma}_k^\star)$ is a vector that is not an $M$-estimator, we first introduce the spatial median aggregation of general vector-valued estimators.

 \subsection{ Spatial Median Aggregation of General Vector-Valued Estimators}\label{sec:spatial_med_of_vector_value}

In this subsection, we consider aggregating general  estimators $\widehat{\bmeta}_k^\star \in \mathbb{R}^{d}$ for $k = 1, \cdots, K$. In particular, their uncontaminated versions, $\widehat{\bmeta}_k$, are not necessarily $M$-estimators, though we assume they satisfy
  \be\sqrt{n_k}\lsk\widehat{\bmeta}_k-\bmeta^{(0)}\rsk\stackrel{d}{\longrightarrow} \mathcal{N}\lsk\boldsymbol{0}_d,\boldsymbol{C}\rsk, \textrm{ as }n_k\to\infty,\label{local_spatial_assume}\ee 
for $k=1,\cdots,K$, where  $\bmeta^{(0)}$ is the true parameter vector of $\bmeta$, and 
 $\boldsymbol{C}\in\mathbb{R}^{d\times d}$ is the asymptotic variance-covariance matrix of $\widehat{\bmeta}_k$. In this setting, $\vech(\widehat{\bmSigma}_k)$ above is a special case of $\widehat{\bmeta}_k$ because, although $\vech(\widehat{\bmSigma}_k)$ is not an $M$-estimator, we can establish 
 its asymptotic normality based on (\ref{local_estimate_sigma}).
 
Next, we adapt the spatial median estimation introduced in \cite{brown1983statistical} to propose  a new robust aggregation  of $\widehat{\bmeta}_k^\star$. Specifically, the spatial median aggregation of  $\widehat{\bmeta}_k^\star$, denoted by $\widehat{\bmeta}_S$, is obtained by minimizing the objective function 
\be
\mL^\star(\bmeta) =\sum_{k=1}^{K} \laak\sqrt{n_k}\lsk\widehat\bmeta_k^\star-\bmeta\rsk\raak_{2},\label{spatial_median} 
 \ee
where $\sqrt{n_k}$ is included in the objective function because it is the rate of convergence in the asymptotic normality 
 (\ref{local_spatial_assume}), and recall that $\|\cdot\|_2$ is  the vector 2-norm. Revisit  $\mathscr{E}_{\bmtheta}$ and $\mathscr{E}_{\bmSigma}$, and similarly define $\mathscr{E}_{\bmeta}=\{k:\widehat{\bmeta}_k^\star\neq \widehat{\bmeta}_k\}\subset\{1,\cdots,K\}$.  
 In Theorem \ref{tm3} below, we demonstrate that the spatial median aggregation $\widehat{\bmeta}_S$ is $\sqrt{N}$-consistent and is resistant to contamination.

\begin{theorem}\label{tm3} Suppose the Asymptotic Setting (\ref{eq:asysetting})  is satisfied. In addition, assume that  
(\ref{local_spatial_assume}) holds 
uniformly for $k=1,\cdots,K$, and  $\limsup_{n_k} \rmE\{\|\sqrt{n_k}(\widehat{\bmeta}_k - \bmeta^{(0)})\|_2^{-2}\} \leq C_{\bmeta}$ for all $k=1,\cdots,K$, where $C_{\bmeta}$ is a finite positive constant. 
If  
$|\mathscr{E}_{\bmeta}|/\sqrt{K}\to 0$ as $N\to\infty$, then we obtain 
 $\widehat{\bmeta}_S-\bmeta^{(0)}=O_\rmP( N^{-1/2})$.
\end{theorem}


 \noindent Analogous to  Theorem \ref{uni:consist2}, Theorem \ref{tm3} requires the  ``fault tolerance"  condition  $|\mathscr{E}_{\bmeta}|/\sqrt{K}\to 0$, but does not impose any restrictions on how the estimates in   $\{\widehat{\bmeta}_k^\star : k \in \mathscr{E}_\bmeta\}$ are contaminated. However, since Theorem \ref{tm3} achieves only $\sqrt{N}$-consistency and not asymptotic normality, it does not require the condition $n^2/N \to \infty$ as specified in Theorem \ref{uni:consist2}; details can be found in the proof of Theorem \ref{tm3} in  Section \ref{proof:tm} of the supplementary material.

Theorem \ref{tm3} is established for the spatial median aggregation of general vector-valued estimators that satisfy (\ref{local_spatial_assume}), and its conditions differ from  Conditions (C1) -- (C5) required in Theorem \ref{uni:consist2}. An additional moment condition  
$\limsup_{n_k} \rmE\{\|\sqrt{n_k}(\widehat{\bmeta}_k - \bmeta^{(0)})\|_2^{-2}\} \leq C_{\bmeta}$ is imposed in Theorem \ref{tm3}. We outline the necessity of imposing such a condition. We first denote the uncontaminated version of $\mL^\star(\bmeta)$ in (\ref{spatial_median}) by $\mL(\bmeta)=\sum_{k=1}^{K} \|\sqrt{n_k}(\widehat\bmeta_k-\bmeta)\|_{2}$. In the proof of Theorem \ref{tm3} in Section \ref{proof:tm} of the supplementary material, we need to expand  $\mL(\bmeta)$ at $\bmeta^{(0)}$ as
\bea
\mL\lsk\bmeta\rsk&=&\mL\lsk\bmeta^{(0)}\rsk-\lbk \sum_{k=1}^{K}\frac{\sqrt{n_k}\lsk\widehat{\bmeta}_k-\bmeta^{(0)}\rsk^\top}{\laak\widehat{\bmeta}_k-\bmeta^{(0)}\raak_2}\rbk\lsk\bmeta-\bmeta^{(0)}\rsk\nonumber\\&&+\lsk\bmeta-\bmeta^{(0)}\rsk^\top\lsk\sum_{k=1}^{K}{{n_k}}\bmS_{n_k}\rsk \lsk\bmeta-\bmeta^{(0)}\rsk+\mathscr{R}_N,\nonumber
\eea
where 
\[\bmS_{n_k}=\frac{1}{2\laak\sqrt{n_k}(\widehat{\bmeta}_k-\bmeta^{(0)})\raak_2}\lbk \bmI_d-\frac{\sqrt{n_k}\lsk\widehat{\bmeta}_k-\bmeta^{(0)}\rsk\sqrt{n_k}\lsk\widehat{\bmeta}_k-\bmeta^{(0)}\rsk^\top}{\laak\sqrt{n_k}(\widehat{\bmeta}_k-\bmeta^{(0)})\raak_2^2}\rbk,
\] 
and the remainder term $\mathscr{R}_N$ satisfies  $|\mathscr{R}_N|\leq C_{\bmeta}^\prime\sum_{k=1}^{K}(\sqrt{n_k}{\|\bmeta-\bmeta^{(0)}\|_2^{2+\delta}}/{\|\widehat{\bmeta}_k-\bmeta^{(0)}\|_2^{1+\delta}})$ for some  finite positive constants $\delta$ and $C_{\bmeta}^\prime$  by (A3) in \cite{mottonen2010asymptotic}. Employing (\ref{local_spatial_assume}) and the continuous mapping theorem, we have
\[
\bmS_{n_k} \stackrel{d}\longrightarrow \bmS\textrm{ as }n_k\to\infty, \textrm{ where }  
\bmS= \frac{1}{2\|\bmcalZ\|_2}\lsk \bmI_d-\frac{{\bmcalZ\bmcalZ^\top}}{{\|\bmcalZ\|_2^2}}\rsk\in\mathbb{R}^{d\times d}\textrm{ and }\bmcalZ\sim \mathcal{N}\lsk\boldsymbol{0}_d,\boldsymbol{C}\rsk.
\]
To demonstrate the $\sqrt{N}$-consistency of $\widehat{\bmeta}_S$ in Theorem \ref{tm3}, we need to show 
$N^{-1}\sum_{k=1}^{K}{{n_k}}\bmS_{n_k}$ converges to a positive definite matrix in probability under the Asymptotic Setting (\ref{eq:asysetting}). Specifically, the proof of Theorem \ref{tm3} verifies: (I) $\rmE(N^{-1}\sum_{k=1}^{K}{{n_k}}\bmS_{n_k})\to\rmE(\bmS)$, where $\rmE(\bmS)$ is positive definite by \citet[p.185]{mottonen2010asymptotic}; and (II) $\var\{N^{-1}\sum_{k=1}^{K}{{n_k}}\vec(\bmS_{n_k})\}=o(1)$, where $\vec(\cdot)$ refers  to  the vectorization of any generic matrix. It turns out that the justification of both (I) and (II) requires the condition  
$\limsup_{n_k} \rmE\{\|\sqrt{n_k}(\widehat{\bmeta}_k - \bmeta^{(0)})\|_2^{-2}\} \leq C_{\bmeta}$. For example, this condition implies $\bmS_{n_k}$ is uniformly integrable. This, in conjunction with $\bmS_{n_k} \stackrel{d}\longrightarrow \bmS$, leads to $\rmE \bmS_{n_k}\to \rmE \bmS$ by Example 2.21 in \citet{van2000asymptotic}, and then further yields (I) under the Asymptotic Setting (\ref{eq:asysetting}). See the detailed justification of both (I) and (II) in the proof of Theorem \ref{tm3} in Section  \ref{proof:tm} of the supplementary material. 

By Theorem \ref{tm3}, the spatial median aggregation is $\sqrt N$-consistent and is robust to contamination. Accordingly, we can consider using this approach (instead of Huber aggregation) to aggregate local $M$-estimators. However, spatial median aggregation is less asymptotically efficient than Huber aggregation, so it is less attractive for aggregating local $M$-estimators.  On the other hand, as we  
demonstrate in the following subsection, spatial median aggregation of local variance estimators has the huge advantage that the aggregated estimator is guaranteed to be positive definite.



\subsection{ Spatial Median Aggregation of Local Variance Estimators}

\label{sec:3.2}

The spatial median aggregation of local variance estimators $\widehat{ \bmSigma}_k^\star$ can be obtained by first obtaining ${\widehat{\bmeta}}_S=\arg \min_{\bmeta}\big[\sum_{k=1}^K \|\sqrt{n_k}\{\vech(\widehat{ \bmSigma}_k^\star)-\bmeta\}\|_{2}\big]$ and then computing $\widehat\bmSigma_S=\vech^{-1}({\widehat{\bmeta}}_S)$, where $\vech^{-1}(\cdot)$ is the inverse of $\vech(\cdot)$ such that $\vech^{-1}({\widehat{\bmeta}}_S)$ is a $p\times p$ symmetric matrix. In the following theorem, we demonstrate the positive definiteness of $\widehat\bmSigma_S=\vech^{-1}({\widehat{\bmeta}}_S)$.

\begin{theorem}\label{tm4:pd}
If $\widehat{ \bmSigma}_k^\star$ are positive definite for all $k=1,\cdots,K$, then
$\widehat{\bmSigma}_S$ is positive definite. 
\end{theorem}



The spatial median ${\widehat{\bmeta}}_S=\arg \min_{\bmeta}\big[\sum_{k=1}^K \|\sqrt{n_k}\{\vech(\widehat{ \bmSigma}_k^\star)-\bmeta\}\|_{2}\big]$ considered in this paper is an extension of classical spatial median $\arg \min_{\bmeta}\big[\sum_{k=1}^K \|\{\vech$ $(\widehat{ \bmSigma}_k^\star)-\bmeta\}\|_{2}\big]$, which  is located  within the convex hull of $\vech(\widehat{ \bmSigma}_k^\star)$ for $k=1,\cdots,K$ (see, e.g., Remark 2.1 in \citealp{minsker2015geometric}). In the proof of Theorem \ref{tm4:pd} in Section \ref{proof:tm} of the supplementary material, we show a similar property for ${\widehat{\bmeta}}_S$, which further leads to Theorem \ref{tm4:pd}.

 Recall that the aim of aggregating $\widehat{\boldsymbol{\Sigma}}_k^\star$ is to obtain a positive definite and robust estimator $\widehat{\boldsymbol{\Sigma}}$ that satisfies the condition $\widehat{\boldsymbol{\Sigma}} - \bmSigma = O_\rmP(1/\sqrt{n})$. This ensures that $\widehat{\boldsymbol{\Sigma}}$ can be reliably used in the estimating equations (\ref{robust:estimating eq2}) as well as for conducting statistical inference on $\bmtheta$.  Theorem \ref{tm3}  shows that $\widehat\bmSigma_S$ is robust and satisfies the fast convergence rate $\widehat\bmSigma_S - \bmSigma = O_\rmP(1/\sqrt{N})$ without imposing restrictions on how the estimates in   $\{\widehat{\boldsymbol{\Sigma}}_k^\star : k \in \mathscr{E}_{\bmSigma}\}$ are contaminated, while Theorem \ref{tm4:pd} guarantees the positive definiteness. However, in practice, the contamination of $\widehat{\bmSigma}_k^\star$ for $k \in \mathscr{E}_{\bmSigma}$ can be arbitrary, and thus these matrices may not be positive definite, leading to a failure to satisfy the condition in Theorem \ref{tm4:pd}. Note that Theorem \ref{tm3} does not impose any restrictions on the values of $\widehat{\boldsymbol{\Sigma}}_k^\star$  for $k \in \mathscr{E}_{\bmSigma}$ as long as $|\mathscr{E}_{\bmSigma}|/\sqrt{K}\to 0$. Accordingly, in applications,  the following additional steps can be considered if some $\widehat{\bmSigma}_k^\star$ received at the central processor is not positive definite. First, we consider the symmetrization $[\widehat{ \bmSigma}_k^\star]_{\mathbb{S}}=(\widehat{\bmSigma}_k^\star +\widehat{\bmSigma}_k^{\star\top})/2$. Then, we obtain the projection of $[\widehat{ \bmSigma}_k^\star]_{\mathbb{S}}$ onto the positive definite space, which is $\widetilde{\bmSigma}_k^\star =\arg\min_{\mathbf{Q}\succeq \epsilon \bmI_d}\|[\widehat{ \bmSigma}_k^\star]_{\mathbb{S}}-\mathbf{Q}\|_F^2=\sum_{j=1}^p  \max \{v_j([\widehat{ \bmSigma}_k^\star]_{\mathbb{S}}), \epsilon\}\bmgamma_j \bmgamma_j^{\top}$, where $\mathbf{Q}\succeq \epsilon \bmI_d$
means that $\mathbf{Q}- \epsilon \bmI_d$ is positive semidefinite, $\|\cdot\|_F$ is the Frobenius norm, $\bmgamma_j$ is the eigenvector corresponding to the $j$-th largest eigenvalue $v_j([\widehat{ \bmSigma}_k^\star]_{\mathbb{S}})$ of $[\widehat{ \bmSigma}_k^\star]_{\mathbb{S}}$, and $\epsilon=10^{-5}$ suggested by \cite{xue2012positive}. Since $\widetilde{\bmSigma}_k^\star$ is guaranteed to be positive definite through the above procedure,  replacing those non-positive definite $\widehat{\bmSigma}_k^\star$ received at the central processor with $\widetilde{\bmSigma}_k^\star$ to obtain the spatial median aggregation $\widehat{\bmSigma}_S$,  ensures the positive definiteness of $\widehat{\bmSigma}_S$ by Theorem \ref{tm4:pd}.

In applications, the uncontaminated local variance estimator $\widehat{\bmSigma}_k$ is guaranteed to be  positive definite based on (\ref{local_estimate_sigma}), so if some $\widehat{\bmSigma}_k^\star$ received at the central processor is not positive definite, then $\widehat{\bmSigma}_k^\star$ is contaminated, or equivalently, $k \in \mathscr{E}_{\bmSigma}$. In the following section, we introduce a  two-step approach to sequentially detect the contaminated $\widehat{\bmtheta}_k^\star$ and $\widehat{\bmSigma}_k^\star$ among different servers $k=1,\cdots,K$.

\section{ Detection of Contaminated Estimates}\label{sec:detect}


In practice, it is of interest to detect the contaminated $M$-estimates $\widehat{\bmtheta}_k^\star$ and variance estimates $\widehat{\bmSigma}_k^\star$ among different servers $k=1,\cdots, K$. If any contaminated $\widehat{\bmtheta}_k^\star$ (or $\widehat{\bmSigma}_k^\star$) is detected for local server $k$, as discussed in the Introduction, several potential sources of contamination can be checked, including but not limited to: (i) data contamination at local server $k$; (ii) malfunction of local server $k$; (iii) intentional contamination of the estimate for data privacy and security purposes; and (iv) communication failure between local server $k$ and the central processor. These checks will facilitate timely maintenance, and ensure optimal and efficient operating conditions for the distributed computing system (see, e.g., \citealp{ji2021statistics}).

Before proposing the two-step approach to sequentially detect the contaminated $\widehat{\bmtheta}_k^\star$ and $\widehat{\bmSigma}_k^\star$, we first introduce the key idea. By (\ref{local assum}), Slutsky's theorem, and the continuous mapping theorem, for each given $k$, if $\widehat{\bmtheta}_k^\star$ is not contaminated, then its Mahalanobis distance (\citealp{chandra1936generalised}) to $\bmtheta^{(0)}$ is
 \ben
d_1(\widehat{\bmtheta}_{k}^\star,\bmtheta^{(0)})=\sqrt{n_k\lsk\widehat{\bmtheta}_{k}^\star- \bmtheta^{(0)}\rsk^\top\widehat{\bmSigma}^{-1}\lsk\widehat{\bmtheta}_{k}^\star- \bmtheta^{(0)}\rsk}\label{FSDdistance}\stackrel{d}\longrightarrow \sqrt{\chi^2_p},\een
where $\widehat{\bmSigma}$ is a consistent estimator of $\bmSigma$ which can be  either $\widehat{\bmSigma}_{k_0}$ defined below (\ref{local_estimate_sigma}) or $\widehat{\bmSigma}_S$ defined in Section \ref{sec:3.2}, and $\chi^2_p$ is the chi-squared distributed random variable with  $p$   
degrees of freedom.  
On the other hand, if both $\widehat{\bmtheta}_k^\star$ and $\widehat{\bmSigma}_k^\star$ are not contaminated, then 
 \ben
d_2(\widehat{\bmtheta}_{k}^\star,\bmtheta^{(0)})=\sqrt{n_k\lsk\widehat{\bmtheta}_{k}^\star- \bmtheta^{(0)}\rsk^\top\widehat{\bmSigma}_k^{\star  -1}\lsk\widehat{\bmtheta}_{k}^\star- \bmtheta^{(0)}\rsk}\label{FSDdistance}\stackrel{d}\longrightarrow \sqrt{\chi^2_p}.\een
However, in real applications, $\bmtheta^{(0)}$ is unknown for both distances. Therefore, we substitute it with the Huber-type aggregation $\widehat{\bmtheta}$, leading to the following result.
\begin{corollary}\label{tm:chi2}
Under the conditions given in Theorem \ref{uni:consist2}, for each given $k$, if $\widehat{\bmtheta}_k^\star$ is not contaminated, we obtain $d_1(\widehat{\bmtheta}_{k}^\star,\widehat{\bmtheta})\stackrel{d}\longrightarrow \sqrt{\chi^2_p}$; if both $\widehat{\bmtheta}_k^\star$ and $\widehat{\bmSigma}_k^\star$ are not contaminated, we have  $ d_2(\widehat{\bmtheta}_{k}^\star,\widehat{\bmtheta})\stackrel{d}\longrightarrow \sqrt{\chi^2_p}$. 
\end{corollary}

Based on this corollary, the two-step approach to sequentially detect the contaminated $\widehat{\bmtheta}_k^\star$ and $\widehat{\bmSigma}_k^\star$ is:
\begin{itemize}
  \item Step 1 (Detection of the contaminated $\widehat{\bmtheta}_k^\star$): If $ d_1(\widehat{\bmtheta}_{k}^\star,\widehat{\bmtheta})>\sqrt{\chi^2_{p,\alpha}}$, then $\widehat{\bmtheta}_{k}^\star$ is deemed to be contaminated, where ${\chi^2_{p,\alpha}}$ is 
 the upper-$\alpha$ quantile of $\chi^2_p$. 
  \item Step 2 (Detection of the contaminated  $\widehat{\bmSigma}_k^\star$):    If $\widehat{\bmtheta}_{k}^\star$ is not deemed to be contaminated and $ d_2(\widehat{\bmtheta}_{k}^\star,\widehat{\bmtheta})>\sqrt{\chi^2_{p,\alpha}}$, then $\widehat{\bmSigma}_k^\star$ is deemed to be contaminated.
\end{itemize}

In practice, $\alpha$ can be 0.05 or 0.1, which corresponds to 95\% or 90\% confidence for the detection, respectively. As far as we are aware, the existing literature on distributed computing has neither employed robust aggregation methods in constructing distances for contamination detection in Step 1 (see, e.g.,  \citealp{ji2010distributed} and \citealp{ji2021statistics}), nor considered contamination detection for variance estimates in Step 2. Specifically, these works used some non-robust estimators, e.g., the  weighted average aggregation $\overline{\bmtheta}$ defined below (\ref{robust:estimating eq2}) to replace $\bmtheta^{(0)}$ in $d_1(\widehat{\bmtheta}_{k}^\star,\bmtheta^{(0)})$. This will cause two issues that do not arise in our approach. First, after the replacement, we cannot develop an asymptotic result for $d_1(\widehat{\bmtheta}_{k}^\star,\overline{\bmtheta})$ like that in Corollary \ref{tm:chi2} because $\overline{\bmtheta}$ lacks robustness and is not a $\sqrt{N}$-consistent estimator of $\bmtheta^{(0)}$ in the presence of contamination.  The latter is crucial for obtaining the result in Corollary \ref{tm:chi2}; see the proof of Corollary \ref{tm:chi2} in Section \ref{proof:tm} of the supplementary material. Second, when one or more contaminated local estimates are present, $d_1(\widehat{\bmtheta}_{k}^\star,\overline{\bmtheta})$ may not be large even if $\widehat{\bmtheta}_{k}^\star$ is contaminated, a phenomenon referred to as a  masking effect in the literature (see \citealp{rousseeuw1990unmasking}). Accordingly, the use of the Huber-type aggregation $\widehat{\bmtheta}$ plays  an important role in  our contamination detection. 

\section{ Simulation Studies}\label{numberstud}

In this section, we conduct simulation studies to investigate the numerical performance of the proposed methods in Sections \ref{methodology} -- \ref{sec:detect}. 
All simulation results, except Tables \ref{tb:log:no:sub} -- \ref{tb:log:omniscient:sub}, are provided in Section \ref{sup:sec:sim:tb} of the supplementary material to save space.

In these studies, the datasets $\boldsymbol{Z}_{i}=(Y_i,\boldsymbol{X}_{i}^{\top})^{\top}$  for $i=1,\cdots,N$ are simulated from logistic and linear regression models, where $Y_i$ is the response variable and $\boldsymbol{X}_{i}=(X_{i 1}, X_{i 2})^{\top}$  is the covariate vector. In particular, for logistic regression,  $Y_{i}$ is binary and generated from a Bernoulli distribution with 
\be\label{eq:logistic}
\rmP(Y_{i}=1 | \boldsymbol{X}_{i};\bmtheta^{(0)})=\frac{\exp (\boldsymbol{X}_{i}^{\top} \bmtheta^{(0)})}{1+\exp (\boldsymbol{X}_{i}^{\top} \bmtheta^{(0)})}.
\ee
For linear regression,  $Y_{i}$ is  generated from 
$Y_{i}=\boldsymbol{X}_{i}^{\top}\bmtheta^{(0)}+\varepsilon_i,$
where $\varepsilon_i$ is independently drawn from $\mathcal{N}(0,1)$. For both models, the vector  $\bmtheta^{(0)}$ is set to be $(\theta_1^{(0)},\theta_2^{(0)})^{\top}=(2,1)^{\top}$, and the covariates $X_{i 1}$ and $X_{i 2}$ are independently generated  from  $\mathcal{N}(0,1)$.

To simulate the environment of a distributed system, we consider that the full dataset $\{\bmZ_1,\cdots,\bmZ_N\}$ is distributed across $k=1,\cdots,K$ servers, and
 each server $k$ possesses a subset of data of size $n_k\equiv n $. Thus, $N = n  K$. In our simulation studies, we consider $K=60$, $80$ and $100$, and $n=5,000$, $10,000$ and $15,000$. 
 To estimate $\bmtheta^{(0)}$, each server $k$ obtains the $M$-estimate $\widehat{\bmtheta}_k$ by maximizing the local criterion function (\ref{localcriterion}). For logistic regression
 we set $m(\bmZ_i;\bmtheta)=Y_i\log\{\rmP(Y_{i}=1 | \boldsymbol{X}_{i};\bmtheta)\}+(1-Y_i)\log\{1-\rmP(Y_{i}=1 | \boldsymbol{X}_{i};\bmtheta)\}$ in (\ref{localcriterion}), and $\widehat{\bmtheta}_k$ is the maximum likelihood estimate. For linear regression,  $m(\bmZ_i;\bmtheta)=-(Y_i-\boldsymbol{X}_{i}^{\top}\bmtheta)^2$ in (\ref{localcriterion}), and  $\widehat{\bmtheta}_k$ is  the least-squares estimate. Subsequently, 
we compare the Huber-type aggregation $\widehat{\bmtheta}$ and the weighted average aggregation $\overline{\bmtheta}$ of the estimates $\{\widehat{\bmtheta}_k^\star:k=1,\cdots,K\}$ received  at the central processor. We consider a contamination-free setting, where $\widehat{\bmtheta}_k^\star=\widehat{\bmtheta}_k$ for all $k=1,\cdots,K$, and five contamination settings, where $\widehat{\bmtheta}_k^\star\neq \widehat{\bmtheta}_k$ for $k=1,\cdots,\lfloor K^{1/4}\rfloor$ and $\widehat{\bmtheta}_k^\star= \widehat{\bmtheta}_k$ for $k=(\lfloor K^{1/4}\rfloor+1),\cdots,K$, with $\lfloor\cdot\rfloor$ denoting the floor function. Accordingly, 
$\mathscr{E}=\{1,\cdots,\lfloor  K^{1/4}\rfloor\}$ represents the set of contaminated $\widehat{\bmtheta}_k^\star$. Among the five contamination settings, three involve direct contamination of the estimates $\widehat{\bmtheta}_k^\star$ for $k \in \mathscr{E}$: (i) Omniscient Contamination $\widehat{\bmtheta}_k^{\star}=(-10^{6},-10^{6})^\top$; (ii) Gaussian Contamination   $\widehat{\bmtheta}_k^{\star}$ is randomly drawn from $\mathcal{N}(\boldsymbol{0}_2,200\bmI_2 )$; and (iii) Bit-flip Contamination $\widehat{\bmtheta}_k^{\star}= -\widehat{\bmtheta}_k$. See \cite{xie2018generalized} and  \cite{tu2021variance} for similar contamination settings. The other two settings involve contamination of the data $\{\bmZ_i : i \in \mathcal{S}_k \}$ on local server $k$ for $k \in \mathscr{E}$, which consequently results in a contaminated estimate $\widehat{\bmtheta}_k^\star$. To save space, the details of these two data contamination settings are given in Section \ref{sup:sec:ASimu1} of the supplementary material.

To obtain the Huber-type aggregation $\widehat\bmtheta$  which solves (\ref{robust:estimating eq2}), we require $\widehat\bmSigma$. Using the robust aggregation method in Section \ref{sec:3.2}, we let $\widehat\bmSigma=\widehat\bmSigma_S$, namely the spatial median aggregation of the local variance estimates  $\widehat{\bmSigma}_k^\star$ for $k=1,\cdots,K$. For simplicity, we set $\widehat{\bmSigma}_k^\star=(\widehat{\bmU}_{k}^\star)^{-1}\widehat{\bmV}_{k}^\star(\widehat{\bmU}_{k}^\star)^{-1}$, where $\widehat{\bmV}_{k}^\star={n_{k}^{-1}}\sum_{i\in\mathcal{S}_{k}}\{ \nabla m(\bmZ_i;\widehat{\bmtheta}_{k}^\star)-
  \nabla \overline{M}_k(\widehat{\bmtheta}_{k}^\star)\}\{\nabla m(\bmZ_i;\widehat{\bmtheta}_{k}^\star) -
  \nabla\overline{M}_k(\widehat{\bmtheta}_{k}^\star)$ $\}^\top$ and $\widehat{\bmU}_{k}^\star=-\nabla^2 \overline{M}_k(\widehat{\bmtheta}_{k}^\star)$. Accordingly, if $\widehat{\bmtheta}_k^\star$ is contaminated, then $\widehat{\bmSigma}_k^\star$ is also contaminated; otherwise, $\widehat{\bmSigma}_k^\star$ is equal to the uncontaminated $\widehat{\bmSigma}_k$ as defined in (\ref{local_estimate_sigma}).

We next report our simulation results which are obtained from 1,000 realizations. 
Table \ref{tb:log:no:sub} presents the results for the logistic regression model under the contamination-free setting.  Specifically, we present the averaged bias (BIAS), standard deviation (SD), and averaged standard error (ASE) of the Huber-type aggregation $\widehat{\bmtheta}=(\widehat\theta_1,\widehat\theta_2)^\top$ with the tuning constant set to $c=1.345$ in (\ref{robust:estimating eq2}), and the weighted average aggregation $\overline{\bmtheta}=(\overline\theta_1,\overline\theta_2)^\top$. In addition, Table \ref{tb:log:no:sub} reports the relative efficiency (RE) for the Huber-type aggregation $\widehat{\bmtheta}$, and the empirical coverage probability (CP)  for the 95\% confidence intervals constructed from $\widehat{\bmtheta}$ and $\overline{\bmtheta}$ and their associated asymptotic distributions. To save space, Table \ref{tb:log:no:sub}  only presents the results for  $n=5,000$ and $15,000$, while  Table \ref{tb:log:no:1.345}  reports the results for  $n=10,000$. Moreover, Table \ref{tb:log:no:0.9818}  presents the results for $\widehat{\bmtheta}$ with setting $c=0.9818$.

The precise definitions of the criteria in Tables \ref{tb:log:no:sub}, \ref{tb:log:no:1.345} and \ref{tb:log:no:0.9818} are provided below. 
Given $r=1,\cdots,1,000$ realizations of the  Huber-type aggregation $\widehat{\bmtheta}^{[r]}=({\widehat{\theta}}_1^{[r]},{\widehat{\theta}}_2^{[r]})^\top$ and its associated standard errors $(\textrm{SE}_1^{[r]},\textrm{SE}_2^{[r]})^\top$ in our simulation studies, the BIAS of $\widehat\theta_j$ is  $1000^{-1}\sum_{r=1}^{1000}( {\widehat{\theta}}_{j}^{[r]}-{\theta}_{j}^{(0)})$,  the SD of $\widehat\theta_j$  is $\{1000^{-1}\sum_{r=1}^{1000}$ $({\widehat{\theta}}_{j}^{[r]}-1000^{-1}{\sum_{r=1}^{1000}{\widehat{\theta}}_{j}^{[r]}})^2\}^{1/2}$,  and 
 the ASE of $\widehat\theta_j$ is $ 1000^{-1}\sum_{r=1}^{1000} \textrm{SE}_j^{[r]}$,  where 
 \be\label{eq:SE1}
\textrm{SE}_j=\{N^{-1}\tau_c^{-1}(\widehat{\boldsymbol{\Sigma}}_S)_{jj}\}^{1/2},
\ee
with 
 $(\widehat{\boldsymbol{\Sigma}}_S)_{jj}$ denoting the $(j,j)$-th element of $\widehat{\boldsymbol{\Sigma}}_S$ 
 for $j=1,2$. In addition, 
 the  CP is  $1000^{-1}\sum_{r=1}^{1000}$ $\mathbbm{1}{\big\{ \theta_{j}^{(0)}  \in [\widehat{\theta}}_{j}^{[r]}-1.96\times  \textrm{SE}_j^{[r]}, {\widehat{\theta}}_{j}^{[r]}+1.96 \times \textrm{SE}_j^{[r]}]\big\}$, where $\mathbbm{1}{\{\cdot\}}$ is the  indicator function defined below (\ref{robust:estimating eq2}). 
 The definitions of the BIAS, SD, ASE and CP for the
weighted average aggregation $\overline{\bmtheta}$ are similar, but use instead as the SE of the weighted average aggregation $\overline\theta_j$,
\be\label{eq:SE2}
\textrm{SE}_j=\{N^{-1} (\overline{\boldsymbol{\Sigma}})_{jj}\}^{1/2},
\ee
where $\overline{\boldsymbol{\Sigma}}=\sum_{k=1}^K(n_k/N)\widehat{\bmSigma}_k^{\star}$ and $\overline\theta_j$ is the $j$-th element of the weighted average aggregation $\overline{\bmtheta}$ for $j=1,\cdots,p$.
 Lastly, we follow \cite{fan1992design} and \cite{de2021review} and define the relative efficiency (RE) as the ratio of the $\textrm{MSE}$ of the non-robust aggregation $\overline{\theta}_j$ to the $\textrm{MSE}$ of the robust aggregation $\widehat\theta_j$, where $\textrm{MSE} = \textrm{BIAS}^2+\textrm{SD}^2$.

Tables \ref{tb:log:no:sub}, \ref{tb:log:no:1.345} and  \ref{tb:log:no:0.9818}
reveal four important findings. (I) The BIAS and SD of $\widehat\bmtheta$ (or $\overline\bmtheta$) generally decrease as $n$ gets larger while $K$ remains the same, or as both $n$ and $K$ increase. However, holding $n$ constant and increasing $K$, we observe a decrease in the SD for both $\widehat\bmtheta$ and $\overline\bmtheta$, but the  BIAS remains of similar order. 
This finding is consistent with the comment after Theorem \ref{uni:consist2} that the aggregation of $K$ estimators in the distributed computing system can reduce the variance to $O\big((nK)^{-1}\big)$, but cannot improve the bias which has an order of $O(n^{-1})$.  (II)  
The ASE of the Huber-type aggregation $\widehat\bmtheta$ is close to the SD across all settings, which confirms that  $\tau_c^{-1}\widehat{\bmSigma}_S$ obtained by spatial median aggregation, is a consistent estimator of the asymptotic variance-covariance matrix $\tau_c^{-1}{\bmSigma}$ in Theorem \ref{uni:consist2}. 
(III) The  CP of $\widehat\bmtheta$ is close to the nominal 95\% level across all settings. This means that inference based on the asymptotic normality of  Theorem \ref{uni:consist2} is valid. 
(IV) The RE of  the Huber-type aggregation $\widehat\bmtheta$ is consistent with  the theoretical ARE $\tau_c$ from Theorem \ref{uni:consist2}, where $\tau_c=95.0\%$ for $c=1.345$ and $\tau_c=90.0\%$ for $c=0.9818$.

\begin{table}[htbp!]
\renewcommand{\arraystretch}{0.9}
\setlength{\tabcolsep}{4pt}
\caption{Simulation results for logistic regression under the contamination-free setting. The tuning constant $c$ for obtaining  $\widehat{\bmtheta}$ is $1.345$.  The BIAS, SD and ASE in this table are 100 times  their actual values.}
\label{tb:log:no:sub}
\vspace{-20pt}
\begin{center}
\scalebox{0.8}{
\begin{tabular}{c|c|rr|rr|rr|rr|rr|rrrrrrrrrr|rr|rr}

   \hline

 & \multicolumn{1}{c|}{} & \multicolumn{4}{c|}{$K=60$}  & \multicolumn{4}{c|}{$K=80$} & \multicolumn{4}{c}{$K=100$}\\
 \cline{3-4}\cline{5-6}\cline{7-8}\cline{9-10}\cline{11-12}\cline{13-14}\cline{15-16}\cline{17-18}\cline{19-20}\cline{21-22}\cline{23-24}\cline{25-26}\noalign{\vspace{0.25ex}} 
 $n$ &         &
\multicolumn{2}{c|}{$\widehat{\bmtheta}$}   & \multicolumn{2}{c|}{$\overline{\bmtheta}$}         &
\multicolumn{2}{c|}{$\widehat{\bmtheta}$}   & \multicolumn{2}{c|}{$\overline{\bmtheta}$}          &
\multicolumn{2}{c|}{$\widehat{\bmtheta}$}   & \multicolumn{2}{c}{$\overline{\bmtheta}$}   
\\
\cline{3-4}\cline{5-6}\cline{7-8}\cline{9-10}\cline{11-12}\cline{13-14}\cline{15-16}\cline{17-18}\cline{19-20}\cline{21-22}\cline{23-24}\cline{25-26}
 & \multicolumn{1}{c|}{}&  \multicolumn{1}{c}{${\theta}_1$}& \multicolumn{1}{c|}{${\theta}_2$} & \multicolumn{1}{c}{${\theta}_1$} & 
 \multicolumn{1}{c|}{${\theta}_2$}  
 & \multicolumn{1}{c}{${\theta}_1$} & \multicolumn{1}{c|}{${\theta}_2$} & \multicolumn{1}{c}{${\theta}_1$} & \multicolumn{1}{c|}{${\theta}_2$} & \multicolumn{1}{c}{${\theta}_1$} & \multicolumn{1}{c|}{${\theta}_2$} & \multicolumn{1}{c}{${\theta}_1$} & \multicolumn{1}{c}{${\theta}_2$}&

 \\
  \hline\noalign{\vspace{0.25ex}} 
&BIAS         &0.138&0.105&0.169&0.109  &0.159&0.099&0.163&0.101&0.146&0.100&0.165&0.099\\
 & SD         &0.777&0.569&0.752&0.551  &0.684&0.503&0.663&0.493&0.596&0.445&0.579&0.437\\
5,000
&ASE           &0.783&0.583&0.959&0.714  &0.678&0.505&0.814&0.605&0.607&0.452&0.714&0.531\\
&CP (\%)      &94.6&94.6&95.1&94.3        &93.9&94.2&94.0&93.8  &94.3&95.0&94.2&95.4\\
 &$\text{RE}$ (\%)     &96.1 &95.6    &      &       &94.5 &96.3&   &     &94.6 &96.0          \\
\hline         \noalign{\vspace{0.25ex}}                                                                                                      
&BIAS      &0.059&0.026&0.063&0.037&0.051&0.023&0.064&0.031&0.056&0.022&0.060&0.031\\
 & SD      &0.466&0.343&0.456&0.336&0.393&0.281&0.381&0.272&0.356&0.252&0.347&0.245\\
15,000
&ASE        &0.452&0.337&0.480&0.357&0.392&0.292&0.412&0.307&0.350&0.261&0.366&0.272\\
&CP (\%)   &95.0&94.6&93.7&93.8&94.5&95.1&95.2&95.3&94.2&95.9&94.4&95.7\\
 &$\text{RE}$ (\%)   & 96.0 &96.5&      &                 &95.0& 95.1& &   &95.4& 95.3                      \\
\hline
\end{tabular}}
\end{center}

\vspace{-20pt}

\end{table}

Table \ref{tb:log:omniscient:sub} 
 presents  the BIAS,   SD, ASE and CP   of 
 the Huber-type aggregation $\widehat{\bmtheta}$ with $c=1.345$, and the weighted average aggregation $\overline{\bmtheta}$,    as well as the RE for  $\widehat{\bmtheta}$, for the logistic regression model under Omniscient Contamination. 
 In addition, we follow \cite{leys2018detecting} and report   the average hit rate (HR) in Table \ref{tb:log:omniscient:sub} to assess the contamination detection approach introduced in Section \ref{sec:detect}. Specifically, with 95\% confidence, we estimate the set of contaminated $\widehat\bmtheta_k^\star$ as $\widehat{\mathscr{E}}=\{k:d_1^2(\widehat{\bmtheta}_{k}^\star,\widehat{\bmtheta})>{\chi^2_{2,0.05}}\}\subset\{1,\cdots,K\}$.  We  calculate the correct detection ratio $\textrm{R}=|\widehat{\mathscr{E}}\cap \mathscr{E}|/|\mathscr{E}|$, and then, averaging over the $1,000$ replicates, $\textrm{HR} 
 =1000^{-1}\sum_{r=1}^{1000} \textrm{R}^{[r]}$.
Again, Table \ref{tb:log:omniscient:sub} only presents the results  for $n=5,000$ and $15,000$ to save space, while Table \ref{tb:log:omniscient:1.345} reports results for $n=10,000$. Moreover,   Table \ref{tb:log:omniscient:0.9818} presents the results for $\widehat{\bmtheta}$ with $c=0.9818$.

\begin{table}[htbp!]
\renewcommand{\arraystretch}{0.9}
\setlength{\tabcolsep}{3.7pt}
\caption{Simulation results for logistic regression under Omniscient Contamination. The tuning constant $c$ for obtaining  $\widehat{\bmtheta}$ is $1.345$. The SD and ASE in this table are 100 times  their actual values.  The BIASes  for $\widehat{\bmtheta}$ and $\overline{\bmtheta}$ in this table are 100 and $10^{-3}$ times their actual values, respectively. The RE in this table is $10^{-13}$ times its actual value.}
\label{tb:log:omniscient:sub}
\vspace{-20pt}
\begin{center}
\scalebox{0.8}{
\begin{tabular}{c|c|rr|rr|rr|rr|rr|rrrrrrrrrrrrrr}

 \cline{1-2}\cline{3-4}\cline{5-6}\cline{7-8}\cline{9-10}\cline{11-12}\cline{13-14}

 & \multicolumn{1}{c|}{} & \multicolumn{4}{c|}{$K=60$}  & \multicolumn{4}{c|}{$K=80$} & \multicolumn{4}{c}{$K=100$}\\
 \cline{3-4}\cline{5-6}\cline{7-8}\cline{9-10}\cline{11-12}\cline{13-14}\noalign{\vspace{0.25ex}} 
 $n$ &         &
\multicolumn{2}{c|}{$\widehat{\bmtheta}$}   & \multicolumn{2}{c|}{$\overline{\bmtheta}$}         &
\multicolumn{2}{c|}{$\widehat{\bmtheta}$}   & \multicolumn{2}{c|}{$\overline{\bmtheta}$}          &
\multicolumn{2}{c|}{$\widehat{\bmtheta}$}   & \multicolumn{2}{c}{$\overline{\bmtheta}$}   
\\
\cline{3-4}\cline{5-6}\cline{7-8}\cline{9-10}\cline{11-12}\cline{13-14}
 & \multicolumn{1}{c|}{}&  \multicolumn{1}{c}{${\theta}_1$}& \multicolumn{1}{c|}{${\theta}_2$} & \multicolumn{1}{c}{${\theta}_1$} & 
 \multicolumn{1}{c|}{${\theta}_2$}  
 & \multicolumn{1}{c}{${\theta}_1$} & \multicolumn{1}{c|}{${\theta}_2$} & \multicolumn{1}{c}{${\theta}_1$} & \multicolumn{1}{c|}{${\theta}_2$} & \multicolumn{1}{c}{${\theta}_1$} & \multicolumn{1}{c|}{${\theta}_2$} & \multicolumn{1}{c}{${\theta}_1$} & \multicolumn{1}{c}{${\theta}_2$}&\\

 \cline{1-2}\cline{3-4}\cline{5-6}\cline{7-8}\cline{9-10}\cline{11-12}\cline{13-14}\noalign{\vspace{0.25ex}} 

&BIAS          &-0.153&-0.145&-33.33&-33.33  &-0.197&-0.103&-24.99&-24.99 &-0.142&-0.086&-29.98&-29.98\\
 & SD          &0.780&0.562     &19.48&14.09  &0.671&0.510&14.48&10.50   &0.615&0.451&17.64&12.60\\
&ASE            &0.783&0.583    &2012.7&2043.5 &0.678&0.505&2011.6&2009.4    &0.607&0.452&2033.5&2042.4\\
5,000&CP (\%)       &93.8&94.4&0.0&0.0&94.3        &93.6&0.0&0.0                 &94.7&92.4&0.0&0.0\\
  &$\text{RE}$ (\%)     &  175.8& 329.7&      &           &127.7& 230.8& &                &225.8&   426.8                  \\
   &$\text{RE}^\dag$ (\%)      & 93.6 &95.1&      &           &95.3& 93.5& &                &94.4& 95.8                      \\
&HR (\%)       &\multicolumn{2}{c|}{100.0}&\multicolumn{2}{c|}{} &\multicolumn{2}{c|}{100.0}&\multicolumn{2}{c|}{} &\multicolumn{2}{c|}{100.0}&\\
 \cline{1-2}\cline{3-4}\cline{5-6}\cline{7-8}\cline{9-10}\cline{11-12}\cline{13-14}     \noalign{\vspace{0.25ex}}                                                                                                        
&BIAS       &-0.060&-0.042&-33.33&-33.33 &-0.042&-0.026&-24.99&-24.99&-0.055&-0.022&-29.98&-29.98\\
 & SD       &0.465&0.343 &6.242&4.430&0.395&0.281&4.664&3.305 &0.357&0.253&5.669&3.988
\\
&ASE         &0.452&0.337&2044.4&2032.3&0.392&0.292&2045.3&2021.2&0.350&0.261&2014.3&2016.2\\
15,000&CP (\%)    &93.8&93.7&0.0&0.0&94.0&93.8&0.0&0.0&94.1&95.3&0.0&0.0\\
   &$\text{RE} $  (\%)    & 505.3 &930.3&      &           &396.0&784.6 & &                &689.7& 1395.3                    \\
 &$\text{RE}^\dag$ (\%)   & 96.3 &95.6&      &                 &94.5& 94.1& &   &95.0& 95.3                      \\
&HR (\%)      &\multicolumn{2}{c|}{100.0}&& &\multicolumn{2}{c|}{100.0}&& &\multicolumn{2}{c|}{100.0}&& &&\\
 \cline{1-2}\cline{3-4}\cline{5-6}\cline{7-8}\cline{9-10}\cline{11-12}\cline{13-14}
\end{tabular}}
\end{center}
\vspace{-20pt}
\end{table}

Tables \ref{tb:log:omniscient:sub}, \ref{tb:log:omniscient:1.345} and \ref{tb:log:omniscient:0.9818} reveal four insights. First, under Omniscient Contamination, the performance of the Huber-type aggregation $\widehat{\bmtheta}$ is qualitatively similar to that observed under the contamination-free setting. In particular, the findings (I) -- (III) from the contamination-free setting also apply to $\widehat{\bmtheta}$ in Tables \ref{tb:log:omniscient:sub}, \ref{tb:log:omniscient:1.345}, and \ref{tb:log:omniscient:0.9818}, demonstrating the robustness of the Huber-type aggregation $\widehat{\bmtheta}$. 
This result is not surprising since Theorem \ref{uni:consist2} still holds when $|\mathscr{E}|=\lfloor K^{1/4}\rfloor=o(\sqrt K)$. 
Second, the weighted average aggregation  $\overline{\bmtheta}$ generates seriously distorted estimation results under Omniscient Contamination.  For example, 
the absolute values of BIAS and SD of  the weighted average aggregation $\overline{\bmtheta}$   are much larger than those of $\widehat{\bmtheta}$, 
and so is $\textrm{MSE} = \textrm{BIAS}^2+\textrm{SD}^2$ of $\overline{\bmtheta}$. Accordingly, the RE of the Huber-type aggregation $\widehat\bmtheta$ in these table exceeds 100\%, demonstrating that $\widehat\bmtheta$ achieves much higher efficiency compared to $\overline\bmtheta$ under the contaminated settings. Tables \ref{tb:log:omniscient:sub}, \ref{tb:log:omniscient:1.345}, and \ref{tb:log:omniscient:0.9818} also report $\text{RE}^\dag$, which is defined as the ratio of the $\textrm{MSE}$ of the weighted average aggregation $\overline{\theta}_j$ in the contamination-free setting to the $\textrm{MSE}$ of the Huber-type aggregation $\widehat{\theta}_j$ in the contaminated settings. 
 Consequently, the finding (IV) above from the contamination-free setting also holds for the $\text{RE}^\dag$  in Tables \ref{tb:log:omniscient:sub}, \ref{tb:log:omniscient:1.345}, and \ref{tb:log:omniscient:0.9818}, which confirms 
  $\tau_c$ in Theorem \ref{uni:consist2} under Omniscient Contamination.    
Third, the CP of $\overline{\bmtheta}$ is much lower than the nominal 95\% level; this is due to the fact that both $\overline{\bmtheta}=(\overline\theta_1,\overline\theta_2)^\top$ and its standard error 
$\textrm{SE}_j =\{N^{-1} (\overline{\boldsymbol{\Sigma}})_{jj}\}^{1/2}$ for $j=1,2$ are not resistant to contamination. Fourth, the contamination detection approach introduced in Section \ref{sec:detect} can detect the contaminated estimates with the  HR being 100\% across all of our settings, which shows the usefulness of the approach and Corollary \ref{tm:chi2} in Section \ref{sec:detect}.

 Tables \ref{tb:log:Gauss} -- \ref{tb:log:inY} report the results for $\widehat{\bmtheta}$ and $\overline{\bmtheta}$ under Gaussian Contamination, Bit-flip Contamination, and the two data contamination settings, respectively. In addition, Tables \ref{tb:lr:no} -- \ref{tb:lr:inX} present results for linear regression under the same settings.  
These tables show qualitatively similar findings to those in Tables \ref{tb:log:no:sub}, \ref{tb:log:no:1.345}, and \ref{tb:log:no:0.9818} for the contamination-free setting, and to those in Tables \ref{tb:log:omniscient:sub}, \ref{tb:log:omniscient:1.345}, and \ref{tb:log:omniscient:0.9818} for contaminated settings. 
Additional findings related to these tables are provided in Section \ref{sup:sec:sim:M} of the supplementary material. 

Recall that \citeauthor{yin2018byzantine}'s (\citeyear{yin2018byzantine}) median aggregation and \citeauthor{tu2021variance}'s (\citeyear{tu2021variance}) averaged quantile aggregation are robust aggregation approaches for aggregating local sample means. In Section \ref{sup:sec:sim:mean} and Table \ref{tb:sm:re} of the supplementary material, we demonstrate that, compared to these two approaches, the Huber-type aggregation offers greater flexibility in selecting the preferred AREs in both contamination-free and contaminated settings. In addition, \citet{tu2021variance} suggests using a finite number of quantiles for aggregation in practice, which achieves approximately a 90\% RE (less than its theoretical ARE $3/\pi\approx95.5\%$), as reported in Section \ref{sup:sec:sim:mean} and Table \ref{tb:sm:re}, while the Huber-type aggregation attains  REs of 90\% ($c=0.9818$) and 95\% ($c=1.345$).
In sum, our proposed robust aggregation approaches support our theoretical findings and can be used for empirical applications.

\section{ Real Data Analysis}\label{realdata}


In this section, we employ the proposed methods in Sections \ref{methodology} -- \ref{sec:detect} to analyze the Airline on time data available at \href{https://dataverse.harvard.edu/dataset.xhtml?persistentId=doi:10.7910/DVN/HG7NV7}{https://doi.org/10.7910/DVN/HG7NV7}. This dataset contains comprehensive flight-related information across  $K=29$ U.S. airline companies. 
To simulate the environment of a distributed system, we assume that the data are scattered across $k=1,\cdots,K$ airline companies. Hence,  each airline $k$ is treated as a local server in the system. The list of these airline companies and the local sample size $n_k$ for each airline $k$ 
are provided in Table \ref{tb:airline}  of the supplementary material. In total we have $N=\sum_{k=1}^{K}n_{k}=118,914,458$ observations in the full dataset, and each observation $\bmZ_i=(Y_i,\bmX_i^\top)^\top$ for $i=1,\cdots,N$ represents a flight record that includes details such as the flight's scheduled and actual arrival times, scheduled departure time, and duration. To predict the delay status of a flight, we follow  the approach in \cite{deshpande2012impact} and define the response variable $Y_i=1$ if the actual arrival time is 15 minutes or more after the scheduled arrival time; otherwise, $Y_i=0$. In addition, we construct  the covariate vector  $\bmX_i= (1, X_{i2}, \cdots, X_{ip})^\top$
for  $i = 1, \cdots,N $. First, $X_{i2}$ and $X_{i3}$ are the Duration and Year of the flight, respectively. The rest of the covariates are indicator variables representing three categorical variables Month, 
Day and DepT. Month is a categorical variable with twelve levels, representing the month of the flight, ranging from January to December.
Day is a categorical variable with seven levels, representing the day of the week of the flight, ranging from Monday to Sunday. (The variables Year, Month and Day are all constructed using the flight's scheduled departure time and date.) 
DepT is a categorical variable with four levels: Morning (6:00–11:59), Afternoon (12:00–17:59), Evening (18:00–23:59), and Night (24:00–5:59), representing the period of the flight's scheduled departure time. In total, we have $p=23$ variables in $\bmX_i= (1, X_{i2}, \cdots, X_{ip})^\top$; see Table {\ref{airline:estimate:sub}} for details.
 




 In this distributed system with $k=1,\cdots,K$ servers/airlines, we fit a logistic regression model (\ref{eq:logistic}) in Section \ref{numberstud} relating $Y_i$ to $\bmX_i$. Let $\widehat\bmtheta_k^\star$ denote the maximum likelihood estimate of the regression coefficient vector $\bmtheta$ based on data from each airline $k$; some servers/airlines may have contaminated data, and the superscript $\star$ indicates this potential contamination.
 To aggregate these local estimates $\widehat{\bmtheta}_k^\star$ for $k=1,\cdots,K$, we follow the procedure in Section \ref{numberstud} to obtain the Huber-type aggregation $\widehat{\bmtheta}$ with the tuning constant $c=1.345$ in (\ref{robust:estimating eq2}), and the weighted average aggregation $\overline{\bmtheta}$. We also compute the SEs of $\widehat{\bmtheta}$ and $\overline{\bmtheta}$  based on (\ref{eq:SE1}) and (\ref{eq:SE2}) in Section \ref{numberstud}, respectively.

\begin{table}[htbp!]
\renewcommand{\arraystretch}{0.9}
\caption{The Huber-type aggregation $\widehat\bmtheta$, the weighted average aggregation $\overline{\bmtheta}$, and their SEs for the Airline on time data. The figure in bold indicates that the associated variable is significant at the 5\% level.}
\vspace{-20pt}
\label{airline:estimate:sub}
\begin{center}
\scalebox{0.88}{
\begin{tabular}{rc|r|r|rc|r|r}
   \hline
 \hline
\multicolumn{2}{c|}{Variable }&\multicolumn{1}{c|}{$\widehat{\bmtheta}$ (SE) }& \multicolumn{1}{c|}{$\overline{\bmtheta}$  (SE) }& \multicolumn{2}{c|}{Variable }&\multicolumn{1}{c|}{$\widehat{\bmtheta}$  (SE)  }& \multicolumn{1}{c}{$\overline{\bmtheta}$  (SE) }\\
 \hline
&Intercept                   &\textbf{-1.234} (0.022)& \textbf{-1.158} (0.084)  \\
\hline
&Year	                     &\textbf{-0.018} (0.001) &  \textbf{-0.081} (0.005)  &&February 	                 &\textbf{-0.054} (0.023)&  -0.053 (0.042)                                           \\\cline{1-4}
 &Duration                    & \textbf{0.114} (0.005)&  \textbf{0.108}  (0.007) &&March	                     &\textbf{-0.125} (0.023)& -0.119    (0.087)                                         \\\cline{1-4}
&Tuesday	                 &\textbf{-0.046} (0.018)& -0.044  (0.036) &&April 	                     &\textbf{-0.332} (0.024)& \textbf{-0.313} (0.026)                                            \\
\parbox[t]{2mm}{\multirow{4}{*}{\rotatebox[origin=c]{90}{ Day}}}&Wednesday	& \textbf{0.050} (0.018)&0.046 (0.041)&&May                         &\textbf{-0.335} (0.022)& \textbf{-0.316}   (0.026)                                          \\
 &Thursday 	         & \textbf{0.203} (0.017) &  \textbf{0.190} (0.028)&\parbox[t]{2mm}{\multirow{3}{*}{\rotatebox[origin=c]{90}{ Month}}} &June&\textbf{-0.033} (0.013)& -0.033 (0.034)  \\
&Friday	                     & \textbf{0.252} (0.017)&  \textbf{0.236} (0.018)  &&July	                     &\textbf{-0.094} (0.023)& -0.089     (0.071)                                        \\
&Saturday                    &\textbf{-0.159} (0.019)& \textbf{-0.150} (0.021)&&August      	             &\textbf{-0.141}  (0.023)& -0.134   (0.098)                                         \\
 &Sunday	                     &\textbf{-0.020}  (0.010) & -0.019  (0.019) &&September	                 &\textbf{-0.507}  (0.025)& \textbf{-0.478}  (0.034)                                          \\\cline{1-4}
 \parbox[t]{2mm}{\multirow{3}{*}{\rotatebox[origin=c]{90}{ DepT}}}&Afternoon&\textbf{0.573} (0.050) & \textbf{0.540} (0.083)  &&October                     &\textbf{-0.347}  (0.024)& \textbf{-0.328}  (0.025)                                          \\
 &Evening 		 &\textbf{0.633} (0.031) & -0.007 (0.042) &&November                    &\textbf{-0.248}  (0.024)& \textbf{-0.234}  (0.027)                                          \\
 &Night	                     &\textbf{-0.245} (0.037) & \textbf{-0.232} (0.056)&&December                    & \textbf{0.150}  (0.022) & 0.140  (0.101)                                          \\
  \hline\hline
\end{tabular}}
\end{center}
\vspace{-20pt}
\end{table}

 




Table \ref{airline:estimate:sub} presents  $\widehat{\bmtheta}$ and $\overline{\bmtheta}$, their SEs, and the significance of the variables at the 5\% level. Both $\widehat{\bmtheta}$ and $\overline{\bmtheta}$ exhibit similar patterns across the variables Year, Duration, Day and Month, though the estimated regression coefficients for these variables may vary in magnitude between $\widehat{\bmtheta}$ and $\overline{\bmtheta}$. First, for Year, both $\widehat{\bmtheta}$ and $\overline{\bmtheta}$ indicate a trend of reduced flight delays over time, which may result from technological advancements, infrastructure improvements, or more realistic scheduling practices in recent years. 
Second, for Duration, both $\widehat{\bmtheta}$ and $\overline{\bmtheta}$ reveal that longer flight durations are associated with a higher probability of delays, which aligns with expectations.
 Third, for Day, specific weekdays---such as Wednesday, Thursday, and Friday---are associated with a higher possibility of delays compared to both the reference level (Monday) and weekends, as reflected in their significantly positive  coefficients in $\widehat{\bmtheta}$. 
 These patterns were also reported by \cite{abdel2007detecting}. However, the coefficient of Wednesday in $\overline{\bmtheta}$ is not significant, likely due to the generally larger SE of $\overline{\bmtheta}$ compared to $\widehat{\bmtheta}$ across all variables, as shown in Table \ref{airline:estimate:sub}. This may be attributed to the SE of $\overline{\bmtheta}$ being obtained by (\ref{eq:SE2}), which depends on the weighted average aggregation $\overline{\boldsymbol{\Sigma}} = \sum_{k=1}^K (n_k/N)\widehat{\bmSigma}_k^{\star}$ that is not resistant to contamination. On the contrary, the SE of $\widehat{\bmtheta}$, computed from (\ref{eq:SE1}) uses the spatial median aggregation, which is robust to contamination.
Lastly, $\widehat{\bmtheta}$ associates December with an increased probability of flight delays, potentially due to the holiday demand surge and seasonal winter conditions. In contrast,  the coefficient of December in $\overline{\bmtheta}$ is not significant.




On the other hand, $\widehat{\bmtheta}$ and $\overline{\bmtheta}$ show differences in the effect of DepT. Specifically, the coefficients of Afternoon and Evening in $\widehat{\bmtheta}$ are significantly positive, suggesting these flights are more likely to experience delays compared to the reference level (Morning). This finding aligns with results from \cite{deshpande2012impact} and \cite{arora2020effect}. In contrast, the coefficient of Evening in $\overline{\bmtheta}$ shows a different sign and is not significant, indicating a discrepancy in the estimates between $\widehat{\bmtheta}$ and $\overline{\bmtheta}$.

\begin{figure}[htbp!]
   \centering 
   \includegraphics[scale=0.25]{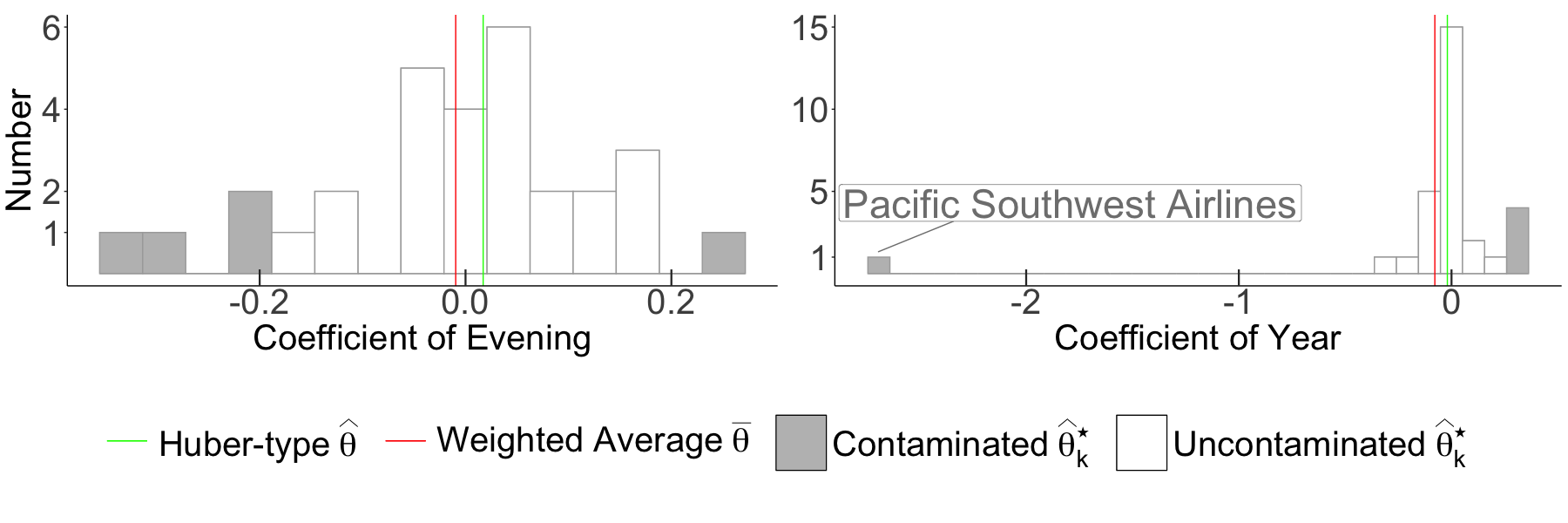}
   \caption{Histograms of the $K=29$ local estimates $\widehat{\theta}_{k,\textrm{Evening}}^\star$ and the $K=29$ local estimates $\widehat{\theta}_{k,\textrm{Year}}^\star$.}
   \label{fig:dist_theta}
   \end{figure}

Next, we apply the contamination detection approach from Section \ref{sec:detect} to identify any contaminated $\widehat{\bmtheta}_k^\star$ among the estimates $\{\widehat{\bmtheta}_k^\star : k = 1, \cdots, K\}$. Specifically, with 95\% confidence, $\widehat{\bmtheta}_k^\star$ is 
 contaminated if $d_1^2(\widehat{\bmtheta}_k^\star,\widehat{\bmtheta}) > {\chi^2_{23,0.05}}$. Among the $K=29$ airlines listed in Table \ref{tb:airline}, five airlines have $\widehat{\bmtheta}_k^\star$ detected as contaminated: Pacific Southwest Airlines, Envoy Air, Piedmont Airlines, Hawaiian Airlines, and Mesa Airlines. 
We focus on interpreting the source of contamination for Pacific Southwest Airlines, as its associated coefficient of Year in $\widehat{\bmtheta}_k^\star$ is negative and much smaller than those of other airlines, as shown in the right panel of Figure \ref{fig:dist_theta}. It is worth noting that the negative coefficient of Year suggests a decreasing probability of delay over time. 
Assuming there are no data quality issues specific to Southwest Airlines, this decreasing probability may reflect improvements to scheduling that may have been introduced following the hijacking incident in 1987 (\citealp{allen1996beyond}) and the integration into U.S. Airways after 1988 (\citealp{miszak1989organizational}). 


 




The detected contaminated $\widehat{\bmtheta}_k^\star$ may also explain the discrepancy observed between the estimates of $\widehat{\bmtheta}$ and $\overline{\bmtheta}$. For instance, the coefficient of Evening in $\overline{\bmtheta}$ has an opposite sign compared to that in $\widehat{\bmtheta}$, and the coefficient of Year in $\overline{\bmtheta}$ is much smaller than in $\widehat{\bmtheta}$; see Table \ref{airline:estimate:sub}. For each local estimate $\widehat{\bmtheta}_k^\star$, $k=1,\cdots,K$, we denote the coefficients of Evening and Year in $\widehat{\bmtheta}_k^\star$  by $\widehat{\theta}_{k,\textrm{Evening}}^\star$ and $\widehat{\theta}_{k,\textrm{Year}}^\star$, respectively.  
Figure \ref{fig:dist_theta} displays histograms of the $K=29$ local estimates $\widehat{\theta}_{k,\textrm{Evening}}^\star$ and the $K=29$ local estimates $\widehat{\theta}_{k,\textrm{Year}}^\star$, respectively.
 The detected contaminated $\widehat{\theta}_{k,\textrm{Evening}}^\star$ and $\widehat{\theta}_{k,\textrm{Year}}^\star$ are also shown in Figure \ref{fig:dist_theta}, along with the estimated coefficients of Evening and Year in $\widehat{\bmtheta}$ and $\overline{\bmtheta}$. It is evident from Figure \ref{fig:dist_theta} that the distortion in the weighted average aggregation $\overline{\bmtheta}$ originates from the five contaminated $\widehat{\bmtheta}_k^\star$. Accordingly, the proposed Huber-type aggregation $\widehat{\bmtheta}$ demonstrates robustness to contamination, and the contamination detection approach from Section \ref{sec:detect} is useful in practice.

\section{ Concluding Remarks}\label{conclude}
In this paper, we propose Huber-type aggregation to robustly aggregate local $M$-estimates across multiple servers in a distributed system, accounting for potential contamination of some local
$M$-estimates. 
Implementation of Huber-type aggregation requires estimation of the asymptotic variance-covariance matrix of the local $M$-estimators,  and we develop the spatial median aggregation of local variance estimators to address this issue.  
This article proves that both Huber-type aggregation and spatial median aggregation achieve the convergence rate of $1/\sqrt{N}$ in the presence of contamination.  We further establish the asymptotic normality of the Huber-type aggregation.  This result is useful for inference and also provides the asymptotic relative efficiency of the Huber-type aggregation. A key advantage of the Huber-type aggregation is that its ARE can be chosen by users. Our theoretical results further lead to a two-step detection approach to identify contamination within the distributed system. 
Numerical studies are conducted to demonstrate the  usefulness of our proposed methods.

To expand the application of the Huber-type aggregation, we identify two future research avenues. 
  First,  the Huber function $\psi_c(\cdot)$ is non-decreasing, while \cite{shevlyakov2008redescending} showed that the Huber  $M$-estimators with this non-decreasing function do not possess finite variance sensitivity. 
  Thus, one can explore replacing $\psi_c(\cdot)$ with the  re-descending Tukey's biweight function \[\varphi_c(u)= \begin{cases}u\lbk1-\lsk\frac{u}{c}\rsk^2\rbk^2 & \text { if }|u| \leq c \\ 0 & \text { otherwise }\end{cases},\]  
 where $\varphi_c(u)$ is non-decreasing near the origin but decreases toward zero as $u$ grows large. Accordingly, using $\varphi_c(\cdot)$ instead of $\psi_c(\cdot)$ can lead to more robust aggregation in the presence of contamination.
Second, in this article, the parameter vector of interest, $\bmtheta$, is assumed to be the same across different local servers. However, in some machine learning literature (see, e.g., \citealp{ma2015partitioning}, \citealp{zhang2018communication} and \citealp{zhang2018communication2}), each local server $k$ can have its own parameter vector of interest $\bmtheta_k$ for $k=1,\cdots,K$, where $\bmtheta_k$ can differ in dimension but may share overlapping parameters. 
It can be valuable to extend the Huber-type aggregation approach to this scenario.
We believe these extensions would further strengthen the usefulness
of  the Huber-type aggregation for distributed data.

\label{sec:co}

\section*{ Appendix: Technical Conditions}

\renewcommand{\theequation}{A.\arabic{equation}}
\setcounter{equation}{0}

\renewcommand{\thesubsection}{A.\arabic{subsection}}

\renewcommand{\thetheorem}{A.\arabic{theorem}}
\setcounter{theorem}{0}

\label{sec:c}
We first introduce the notation used in the technical conditions. 
Let $\|\cdot\|_2$ denote the vector $2$-norm or the matrix $2$-norm. 
  Let $\nabla m( \bmZ;\bmtheta)=\partial  m( \bmZ;\bmtheta)/\partial \bmtheta$, $ \nabla^2 m( \bmZ;\bmtheta)=\partial ^2 m( \bmZ;\bmtheta)/(\partial \bmtheta \partial \bmtheta^\top)$ and $\nabla^{\kappa}m( \bmZ;\bmtheta)=\partial ^\kappa m( \bmZ;\bmtheta)/(\underbrace{\partial \bmtheta^\top\otimes \cdots\otimes \partial \bmtheta^\top}_\kappa)$ for any integer $\kappa\geq 3$. We also denote $\nabla^{\kappa}m( \bmZ;\bmtheta)\big{|}_{\bmtheta=\bmtheta^{(0)}}$  by $\nabla^{\kappa}m( \bmZ;\bmtheta^{(0)})$ for concision.
We next introduce the following technical conditions.

 (C1) 
Assume that the true parameter vector $\bmtheta^{(0)}$ is an interior point of a compact parameter space $\bmTheta\subset\mathbb{R}^p$ and  $\bmtheta^{(0)}$ is the unique maximizer of  $\rmE \{m(\bmZ;\bmtheta)\}	$.  

 (C2) Assume that  $\bmz\mapsto m( \bmz;\bmtheta)$ is measurable  given any $\bmtheta\in\bmTheta$, and $\bmtheta\mapsto m( \bmz;\bmtheta)$ is four times continuously differentiable in $\bmTheta$  for $P_{\bmZ}$-almost every $\bmz$, where $P_{\bmZ}=\rmP\circ {}\bmZ^{-1}$ is the measure induced by the random vector $\bmZ$.
 
 (C3) Assume that $\bmV=\var\{
\nabla m(\bmZ;\bmtheta^{(0)})\}$ and $\bmU=-\rmE\{\nabla^2m(\bmZ;\bmtheta^{(0)})\}$ are finite and  positive definite.

 (C4) Assume that $\rmE \sup_{\bmtheta\in \bmTheta} \|\nabla^{\kappa} m\lsk \bmZ;\bmtheta\rsk\|_2^2<\infty,$  for $\kappa=1,2,3,4$.

 (C5) Assume that $\rmE \|\nabla m( \bmZ;\bmtheta^{(0)})\|_2^4<\infty$.

\noindent Remarks on these conditions are provided in Section \ref{sup:sec:cond_remarks} of the supplementary material.

\bibliographystyle{apalike}

\setstretch{1.48} 
\setlength{\bibsep}{4pt plus 0.3ex}
 \bibliography{reference}

\end{document}